\documentclass[a4paper,fleqn]{cas-dc}

\usepackage[authoryear,longnamesfirst]{natbib}
\usepackage{tabularx}
\usepackage{array}
\usepackage{graphicx}
\usepackage{placeins}
\usepackage{float}

\def\tsc#1{\csdef{#1}{\textsc{\lowercase{#1}}\xspace}}
\tsc{WGM}
\tsc{QE}
\tsc{EP}
\tsc{PMS}
\tsc{BEC}
\tsc{DE}
\begin{document}
\gdef\lastpage{28}
\let\WriteBookmarks\relax
\def\floatpagepagefraction{1}
\def\textpagefraction{.001}

\shorttitle{VeriX-Anon: A Multi-Layered Framework for Mathematically Verifiable Outsourced Target-Driven Data Anonymization}


\title [mode = title]{VeriX-Anon: A Multi-Layered Framework for Mathematically Verifiable Outsourced Target-Driven Data Anonymization}                      
\tnotemark[1,2]

%

\author[1]{Miit Daga}[
    orcid=0009-0005-4629-458X
]
\ead{miit.daga2022@vitstudent.ac.in}

\credit{Conceptualization, Methodology, Software, Validation, Formal analysis, Investigation, Writing - Original Draft}

\author[1]{Swarna Priya Ramu}[
    orcid=0000-0002-8287-9690
]
\cormark[1]
\ead{swarnapriya.rm@vit.ac.in}

\credit{Supervision, Writing - Original Draft, Writing - Review \& Editing, Project administration}

\affiliation[1]{organization={School of Computer Science Engineering and Information Systems, Vellore Institute of Technology},
    city={Vellore},
    postcode={632014},
    state={Tamil Nadu},
    country={India}}

\cortext[cor1]{Corresponding author: Swarna Priya Ramu}

\begin{abstract}
Organisations increasingly outsource privacy-sensitive data transformations to cloud providers, yet no practical mechanism lets the data owner verify that the contracted algorithm was faithfully executed. VeriX-Anon is a multi-layered verification framework for outsourced Target-Driven k-anonymization combining three orthogonal mechanisms: deterministic verification via Merkle-style hashing of an Authenticated Decision Tree, probabilistic verification via Boundary Sentinels and exact-duplicate Twins with cryptographic identifiers, and utility-based verification via Explainable AI fingerprinting that compares SHAP value distributions before and after anonymization using the Wasserstein distance. Across seven cross-domain datasets and four cloud profiles (28 scenarios), against Lazy (drops records), Dumb (fake hash), and Approximate (valid hash) adversaries, VeriX-Anon detects 25 of 28 deviations under a fixed threshold and 27 of 28 once the threshold is calibrated per dataset, with no false alarms. No single layer achieved this alone. The XAI layer was the only mechanism that caught the Approximate adversary, succeeding on six of seven datasets and missing only a high-dimensional case where honest generalization shifts SHAP as much as the attack. Target-Driven anonymization preserved significantly more utility than blind splitting, with mean F1 gaps of $+0.058$ to $+0.362$ and Wilcoxon $p \le 0.001$ on six of seven datasets. Client-side verification completes under one second at one million rows. The threat model covers three empirically evaluated profiles and one theoretical Informed Attacker unable to defeat the cryptographic salt. Sentinel evasion probability ranges from near-zero to 0.82 for the most imbalanced data, which the twin layer offsets in every scenario.
\end{abstract}



\begin{keywords}
k-anonymity \sep verifiable computation \sep intelligent auditing \sep explainable AI \sep Merkle trees \sep SHAP
\end{keywords}

\maketitle

\section{Introduction}
\label{sec:introduction}

The global data centre outsourcing market was valued at USD~127.8 billion in 2024 and is projected to reach USD~157.1 billion by 2030 (\cite{researchandmarkets2025dc}). Organisations across healthcare, finance, and government routinely transfer sensitive datasets to cloud providers for processing, analytics, and anonymization. The economic logic is clear: outsourcing avoids the capital cost of on-premise infrastructure and gives access to elastic compute that small data owners cannot replicate internally.

But outsourcing creates a trust problem. The data owner hands over records containing personally identifiable information (PII) and expects the cloud to apply a specific privacy algorithm, for example, $k$-anonymity via Target-Driven decision tree partitioning. The cloud returns an anonymized dataset, a tree structure, and (at best) a hash. The data owner has no mechanism to verify that the cloud actually ran the contracted algorithm rather than a cheaper shortcut. Yet the data owner, not the cloud, bears the regulatory and reputational cost if the anonymization was performed incorrectly or not at all.

This is not a hypothetical risk. The 2024 IBM Cost of a Data Breach Report puts the average breach cost at USD~4.88 million, a 10\% year-over-year increase and the highest ever recorded (\cite{ibm2024breach}). Healthcare breaches average USD~9.77 million. GDPR fines have exceeded EUR~5.88 billion in aggregate since 2018 (\cite{dlapiper2025gdpr}), and a single outsourcing firm (Capita plc) was fined GBP~14 million in 2025 after a ransomware breach exposed 6.6 million records, affecting 325 of the over 600 pension fund clients it served~(\cite{ico2025capita}). A cloud provider that drops records to save compute or substitutes a faster but utility-destroying algorithm exposes the data owner to penalties of this magnitude.

The need for $k$-anonymity itself is well-established. Sweeney~(\cite{sweeney2002kanonymity}) demonstrated in 1997 that 87\% of the U.S. population can be uniquely identified from just three quasi-identifiers (ZIP code, birth date, sex), and used a \$20 voter registration list to re-identify the Massachusetts governor's medical records. Two decades of follow-up work confirmed the fragility of naive anonymization: Narayanan and Shmatikov~(\cite{narayanan2008netflix}) de-anonymized Netflix users from movie ratings, and de~Montjoye et~al.~(\cite{demontjoye2015credit}) showed that 90\% of individuals in a credit card dataset could be re-identified from just four transactions. $k$-Anonymity addresses this by grouping records so that every individual is indistinguishable from at least $k-1$ others on the quasi-identifier attributes. Target-Driven anonymization~(\cite{friedman2010kdtree}) takes this further: instead of partitioning data blindly, it builds a decision tree that splits on the target variable, preserving predictive utility within each equivalence class. The result is an anonymized dataset that remains useful for downstream classification, not just a privacy-compliant but analytically useless table.

The missing piece is verification. The entire literature on $k$-anonymity focuses on how to perform anonymization correctly. Almost nothing addresses what happens after the data owner outsources the task and receives the result. To our knowledge, no existing framework simultaneously verifies (a) that the correct algorithm was used (structural correctness), (b) that all records were processed (data completeness), and (c) that predictive utility was preserved (algorithmic integrity). Three bodies of prior work are relevant, and none fills this gap on its own.

General-purpose verifiable computation frameworks (~\cite{gennaro2010non}, ~\cite{parno2016pinocchio}) are theoretically sound but impractical for large anonymization trees: the proof generation overhead makes them orders of magnitude more expensive than the original computation for tree-structured tasks with tens of thousands of nodes. Trap-based verification (canary records in database auditing) provides data-level checks but no structural or utility verification; as we show in Section~\ref{sec:detection}, trap-only methods miss an entire class of adversary that substitutes random splitting while processing all records. Explainable AI tools like SHAP~(\cite{lundberg2017shap}) have been used for model debugging and data drift detection, but SHAP has not been used as a verification mechanism for outsourced data transformations. The connection between SHAP value distributions and the algorithmic integrity of an anonymization process is, to our knowledge, new. VeriX-Anon can therefore be understood as an intelligent auditing system: it combines machine learning Random Forest~(RF) boundary detection, Explainable AI~(XAI) via SHAP-based utility fingerprinting, and cryptographic hashing into a single automated verification pipeline that replaces manual or re-execution-based auditing. We note that using SHAP or similar feature attribution methods for detecting data drift or model drift is an active area of research (e.g.,~\cite{chaudhury2024explainable}). Our contribution is not the observation that SHAP distributions shift under data perturbation, which is well established, but the specific application of this shift as a verification signal within a cryptographic auditing pipeline for outsourced anonymization.

The integration of these three mechanisms is not merely additive. As we demonstrate empirically in Section~\ref{sec:detection} (Table~\ref{tab:detection_full}), no single layer detects every deviation on its own. Under the per-dataset calibrated threshold, hash verification (Layer~1) is correct in 14 of 28 scenarios (every honest and every Dumb cloud, but no Lazy or Approximate cloud), trap verification (Layers~2a and~2b together) is also correct in 14 of 28 (every honest and every Lazy cloud, but no Dumb or Approximate cloud), and XAI verification (Layer~3) is correct in 23 of 28 (every honest cloud, four of seven Lazy, six of seven Dumb, and six of seven Approximate). Each layer covers a failure mode the others miss, and only their combination reaches 27 of 28.

This paper presents VeriX-Anon, a multi-layered verification framework that allows a data owner to mathematically audit whether a cloud provider correctly performed Target-Driven $k$-anonymization on outsourced data. The specific contributions are:

\begin{enumerate}
\item A tri-layer intelligent verification architecture combining
deterministic verification (Merkle-style SHA-256 hashing of the
authenticated decision tree), probabilistic verification (Boundary
Sentinels targeting the Random Forest decision boundary at
$P \in [0.45, 0.55]$, plus exact-duplicate Twins), and AI-driven
utility verification (Explainable AI fingerprinting via the Wasserstein distance of SHAP value distributions), forming an integrated expert auditing system.

\item A Boundary Sentinel generation technique that exploits the uncertainty region of a Random Forest classifier to produce synthetic records maximally sensitive to changes in the splitting logic. Unlike generic canary records, these sentinels are placed where algorithm substitution causes the largest displacement.

\item A formal probability analysis of sentinel evasion under data-dropping attacks (Equation~\ref{eq:sentinel_evasion}), with empirical validation showing evasion probabilities ranging from near-zero on balanced datasets (160 sentinels) to 0.82 on the most imbalanced (Diabetes, 4 sentinels under an 89/11 split).

\item Empirical evaluation across seven cross-domain datasets (societal, financial, medical, energy, high-dimensional, and physics) against three adversary profiles (Lazy Cloud: drops 5\% of records; Dumb Cloud: substitutes random splitting with a fake hash; Approximate Cloud: substitutes random splitting with a valid hash), achieving correct detection in 25 of 28 scenarios under a single fixed threshold and 27 of 28 once the Wasserstein threshold is calibrated per dataset. The single remaining evasion is the Approximate adversary on a high-dimensional dataset (Nomao), where honest generalization perturbs the SHAP distribution more than the attack does. No single verification layer achieves full coverage alone.

\item Statistical validation on all seven datasets over an 11-point $k$-sweep: Target-Driven anonymization beats blind splitting in every paired comparison, with Wilcoxon $p \le 0.001$ on six of seven datasets after Benjamini-Hochberg correction, large Cohen's $d$ effect sizes (up to 5.64), and bootstrap 95\% confidence intervals on the mean F1 gap that exclude zero on every dataset.

\item Sub-second client verification overhead: 0.788\,s at $n = 10^6$ rows, with the $O(n/k)$ hash traversal dominated by a fixed $O(1)$ XAI overhead of approximately 0.5\,s.
\end{enumerate}

The remainder of this paper is organised as follows. Section~\ref{sec:related_work} surveys related work on $k$-anonymity, verifiable computation, and XAI-based data quality assurance. Section~\ref{sec:background} defines the technical building blocks that VeriX-Anon relies on. Section~\ref{sec:threat_model} defines the system model, adversary profiles, and security assumptions. Section~\ref{sec:methodology} presents the VeriX-Anon methodology across its four phases. Section~\ref{sec:experiments} reports experimental results, including detection accuracy, scalability, and utility-privacy trade-offs with full statistical analysis. Section~\ref{sec:discussion} discusses limitations, including XAI limitations under severe class imbalance, sentinel density, and epsilon sensitivity. Section~\ref{sec:conclusion} concludes with future directions.

\section{Related Work}
\label{sec:related_work}
Three bodies of prior work intersect with VeriX-Anon: anonymisation algorithms that perform $k$-anonymity (but do not verify it after outsourcing), verifiable computation frameworks that authenticate outsourced results (but were not designed for decision-tree-based anonymisation), and Explainable AI methods that measure data fidelity (but have not been integrated into a cryptographic auditing pipeline). This section reviews each area and identifies the gaps that VeriX-Anon fills.

\subsection{k-Anonymity and Target-Driven Anonymization}

Kesarwani et al. (\cite{kesarwani2021secure}) developed a methodology for secure k-anonymity over encrypted databases using a fully homomorphic encryption framework within a federated cloud architecture. Data partitioning was performed by applying a secure k-means clustering algorithm over encrypted data to group similar tuples prior to generalisation. Although the proposed model achieved data masking and differential privacy without decrypting the dataset, it is to be understood that this is a forward-execution approach without any involvement of post-hoc verification mechanisms, and in this case, the data owner cannot mathematically audit whether the service provider actually executed the correct anonymization algorithm.

De Capitani di Vimercati et al. (\cite{de2024dt}) developed a methodology for target-driven data anonymization using decision trees guided by classification targets. Data generalisation was performed by evaluating candidate splits in a top-down manner and independently anonymizing the resulting leaf nodes to minimize information loss. Although the constructed model partitioned the data to satisfy k-anonymity and l-diversity while maintaining downstream analytic utility, it is to be understood that this is a local transformation approach without any involvement of verifiable outsourced computation, and in this case, delegating this process to an untrusted cloud environment leaves the framework vulnerable to lazy or malicious execution.

Barezzani et al. (\cite{barezzani2025ta_da}) developed a methodology for target-aware data anonymization using a combination of decision tree partitioning and generalized clustering algorithms. The anonymization process was performed by grouping tuples that share similar predictive features and subsequently enforcing privacy constraints on each isolated cluster. Although the stacked anonymization pipeline preserved the predictive features necessary for downstream classification tasks, it is to be understood that this is a classical approach without any involvement of cryptographic auditing structures, and in this case, utilizing it within a multi-controller outsourced environment fails to provide guarantees of structural correctness or completeness.

\subsection{Verifiable Computation and Authenticated Data Structures}

Gennaro et al. (\cite{gennaro2010non}) developed a methodology for non-interactive verifiable computing using Yao's garbled circuits integrated with fully homomorphic encryption. The verification preparation was performed by allowing the client to execute a one-time preprocessing stage that creates a garbled circuit for the target function. Although the theoretical model enabled a computationally weak client to outsource arbitrary functions and verify the returned results, it is to be understood that this is a highly generalized approach without any involvement of specific optimizations for decision tree algorithms, and in this case, the immense proof generation overhead renders it impractical for large-scale dataset anonymization.

Parno et al. (\cite{parno2016pinocchio}) developed a methodology for nearly practical verifiable computation using quadratic arithmetic programs to compile C code into a verifiable cryptographic protocol. The cryptographic translation was performed by mapping arithmetic circuits into a set of polynomials that enable public verification via a short cryptographic proof. Although the Pinocchio system produced verification times of approximately 10 milliseconds, it is to be understood that this is an arithmetic circuit approach without any involvement of embedded probabilistic data traps, and in this case, the worker's computational burden remains too heavy for evaluating massive multidimensional datasets.

Mykletun et al. (\cite{mykletun2006authentication}) developed a methodology for ensuring authentication and integrity in outsourced databases using signature aggregation techniques such as condensed-RSA. The integrity checking was performed by generating tuple-level digital signatures and mathematically aggregating them to provide a unified proof for database query replies. Although the digital signature approach mitigated querier computation and bandwidth overheads, it is to be understood that this is an exact-match verification approach without any involvement of algorithmic auditing, and in this case, it cannot verify complex structural transformations like k-anonymity partitioning.

\begin{table*}[!h]
\centering
\caption{Comparison of verification approaches for outsourced computation. The columns are: Deterministic Verification (a cryptographic guarantee of structural correctness), Probabilistic Traps (embedded records that detect dropped data), Utility Verification (a check that predictive structure is preserved), Cross-Domain Evaluation (tested on more than one data domain), and Practical Complexity (verification cheaper than re-execution, with no special hardware or trusted setup). To the best of our knowledge no prior framework combines the first three for outsourced anonymization: recent surveys of verifiable outsourced learning~(\cite{xing2025zero}) and cloud data-integrity auditing~(\cite{abdul2023state}) report no such combination, and the closest individual scheme~(\cite{de2024query}) adds deterministic and trap checks but no utility verification.}
\label{tab:related_work_comparison}
\resizebox{\textwidth}{!}{ 
\begin{tabular}{|l|c|c|c|c|c|}
\hline
\textbf{Reference} & \textbf{Deterministic} & \textbf{Probabilistic} & \textbf{Utility} & \textbf{Cross-Domain} & \textbf{Practical} \\
& \textbf{Verification} & \textbf{Traps} & \textbf{Verification} & \textbf{Evaluation} & \textbf{Complexity} \\ \hline
Kesarwani et al. (\cite{kesarwani2021secure}) & $\times$ & $\times$ & $\times$ & $\times$ & \checkmark \\ \hline
De Capitani di Vimercati et al. \cite{de2024dt} & $\times$ & $\times$ & $\times$ & $\times$ & \checkmark \\ \hline
Mykletun et al. (\cite{mykletun2006authentication}) & \checkmark & $\times$ & $\times$ & $\times$ & \checkmark \\ \hline
Etemad and K\"{u}p\c{c}\"{u} \cite{etemad2020generic} & \checkmark & $\times$ & $\times$ & $\times$ & \checkmark \\ \hline
Liu et al. (AUDIO) (\cite{liu2012audio}) & Partial & \checkmark & $\times$ & $\times$ & \checkmark \\ \hline
De Capitani di Vimercati et al. (\cite{de2024query}) & \checkmark & \checkmark & $\times$ & $\times$ & \checkmark \\ \hline
Zheng et al. (\cite{zheng2022optimizing}) & Partial & $\times$ & $\times$ & $\times$ & \checkmark \\ \hline
Chaudhury et al. (\cite{chaudhury2024explainable}) & $\times$ & $\times$ & Partial & $\times$ & \checkmark \\ \hline
Gennaro et al. (\cite{gennaro2010non}) & \checkmark (General) & $\times$ & $\times$ & $\times$ & $\times$ \\ \hline
Parno et al. (Pinocchio) (\cite{parno2016pinocchio}) & \checkmark (General) & $\times$ & $\times$ & $\times$ & $\times$ \\ \hline
Setty (Spartan zkSNARK) (\cite{setty2020spartan}) & \checkmark (General) & $\times$ & Partial & $\times$ & $\times$ \\ \hline
Mohassel and Zhang (SecureML) (\cite{mohassel2017secureml}) & $\times$ & $\times$ & $\times$ & \checkmark & $\times$ \\ \hline
Schuster et al. (VC3) (\cite{schuster2015vc3}) & \checkmark & $\times$ & $\times$ & \checkmark & \checkmark \\ \hline
Naive Re-execution Baseline & \checkmark & $\times$ & \checkmark (Trivially) & $\times$ & $\times$ \\ \hline
\textbf{VeriX-Anon (Proposed)} & \checkmark & \checkmark & \checkmark & \checkmark & \checkmark \\ \hline
\end{tabular}
}
\end{table*}
Etemad and K\"{u}p{\c{c}}{\"u} (\cite{etemad2020generic}) developed a methodology for dynamic data outsourcing using implicitly-ordered authenticated data structures coupled with homomorphic verifiable tags. The auditing setup was performed by integrating rank-based skip lists and Merkle trees to support rapid block updates and blockless verification at the cloud server. Although the framework supported dynamic updates and provided strong probabilistic data possession guarantees, it is to be understood that this is a raw data auditing approach without any involvement of explainable AI utility metrics, and in this case, it falls short of verifying the algorithmic correctness of data subjected to clustering or generalization.

Liu et al. (\cite{liu2012audio}) developed a methodology for integrity auditing of outlier-mining-as-a-service systems using the strategic insertion of artificial outlier and non-outlier tuples. The data preparation was performed by randomly sampling the original dataset and constructing specific boundary cases to serve as hidden traps for the semi-honest server. Although the AUDIO framework provided a strong probabilistic guarantee of completeness and correctness for the mining results, it is to be understood that this is a trap-based approach without any involvement of deterministic cryptographic tree structures, and in this case, applying it to target-driven anonymization fails to verify the structural integrity of the generated decision tree.

De Capitani di Vimercati et al. (\cite{de2024query}) developed a methodology for query integrity in smart environments using a hybrid model that combines deterministic authenticated data structures with probabilistic controls like sentinels and twins. The verification process was performed by enriching relational datasets with duplicate records and artificial tuples prior to executing the required SQL operations. Although the combined scheme verified the completeness and correctness of outsourced relational queries, it is to be understood that this is a standard relational algebra approach without any involvement of explainable AI fingerprinting, and in this case, it cannot measure the utility preservation of complex machine learning transformations.

Zheng et al. (\cite{zheng2022optimizing}) developed a methodology for optimizing secure decision tree inference outsourcing using an advanced carry look-ahead adder within an additive secret sharing framework. The secure inference logic was performed by distributing bitwise threshold comparisons across non-colluding servers to eliminate the traditional linear delay of ripple carry adders. Although the system reduced the online inference latency and network communication rounds for cloud servers, it is to be understood that this is a secure inference approach without any involvement of large-scale dataset verification, and in this case, it cannot be scaled to audit the complete construction of a k-anonymity decision tree.

\subsection{Explainable AI for Data Quality Assurance}

Chaudhury et al. (\cite{chaudhury2024explainable}) developed a methodology for explainable artificial intelligence using the Wasserstein distance to quantify model explainability and feature importance. The diagnostic analysis was performed by measuring the optimal transport cost required to morph the probability distribution of key predictive features (such as duration, age, and balance (\cite{moro2011bank})) from a validation set into the distribution of the training data. Although the mathematical formulation evaluated the fidelity of a model and highlighted critical decision boundaries, it is to be understood that this is a post-hoc interpretability approach without any involvement of cryptographic auditing tools, and in this case, it cannot function as a standalone mechanism to verify the execution integrity of an outsourced algorithm.

More recent work places SHAP inside security pipelines for related but distinct ends. An explainable federated-blockchain framework for healthcare~(\cite{bhardwaj2025explainable}) uses SHAP to weight client model updates during aggregation and to log auditable explanations on-chain. There, feature attribution serves transparency and robustness against poisoning, not a check that an outsourced transformation preserved data utility, which is the role SHAP plays in VeriX-Anon.

\subsection{Positioning of VeriX-Anon}

In evaluating the landscape of outsourced computation, it is vital to consider a naive re-execution baseline where the client simply re-runs the full Target-Driven Anonymization locally to verify the correctness of the Cloud's output. The re-execution process is performed by downloading the outsourced results and fully reconstructing the decision tree from the raw feature space (including variables such as duration, age, and balance (\cite{moro2011bank})). Although this naive approach achieves perfect deterministic verification, it is to be understood that this requires $O(n \log n)$ computation to build the full decision tree, and in this case, it defeats the entire purpose of outsourcing the computation to the Cloud. Conversely, VeriX-Anon achieves comprehensive verification in $O(n/k)$ time for the hash traversal alongside an $O(1)$ overhead for the Explainable AI XAI utility check. This renders our approach strictly cheaper and highly practical for resource-constrained clients.

As summarized in Table~\ref{tab:related_work_comparison}, the existing literature addresses isolated facets of outsourced data security, model explainability, and database integrity. Cryptographic frameworks provide rigorous mathematical proofs but incur prohibitive computational costs. Data mining auditing systems utilize artificial traps but lack structural verification and utility analysis. Two recent surveys confirm the gap: a comprehensive survey of zero-knowledge verifiable outsourced machine learning~(\cite{xing2025zero}) catalogues schemes that prove computational correctness at heavy cryptographic cost but none that verify utility, and a survey of data integrity and privacy preservation in the cloud~(\cite{abdul2023state}) covers auditing and $k$-anonymity techniques that check integrity and completeness but not whether predictive structure survived. Furthermore, while the Wasserstein distance effectively measures data fidelity, it has not been integrated into a cryptographic auditing pipeline. VeriX-Anon bridges these critical research gaps by introducing a multi-layered verification framework. By combining $O(n/k)$ authenticated decision tree traversal with probabilistic boundary sentinels and $O(1)$ explainable AI fingerprinting, VeriX-Anon provides a practical, cross-domain solution that avoids the $O(n \log n)$ bottleneck of naive client-side re-execution, ensuring that outsourced anonymization is both mathematically verifiable and utility-preserving.

\section{Background and Fundamentals}
\label{sec:background}

This section defines the technical building blocks that VeriX-Anon relies on. Readers familiar with $k$-anonymity, Merkle trees, and SHAP may skip ahead to Section~\ref{sec:threat_model}.

\subsection{$k$-Anonymity and Quasi-Identifier Generalisation}
\label{sec:bg_kanon}

Sweeney~(\cite{sweeney2002kanonymity}) showed that 87\% of the U.S. population can be uniquely identified from just three attributes: ZIP code, date of birth, and sex. These attributes are called \emph{quasi-identifiers} (QIs), and $k$-anonymity exists to neutralise them. A dataset satisfies $k$-anonymity if every record is indistinguishable from at least $k - 1$ other records on the QI columns. The mechanism is \emph{generalisation}: replace specific QI values with broader ranges until each distinct QI combination appears at least $k$ times. Records sharing the same generalised QI values form an \emph{equivalence class}.

Generalisation comes at a cost. If ages $\{25, 27, 63\}$ are all mapped to $[0, 100]$, the equivalence class is private but analytically worthless. Smaller, tighter classes preserve more information but risk violating the $k$ threshold. Every anonymisation algorithm navigates this tension differently.

\subsection{Target-Driven Decision Tree Anonymisation}
\label{sec:bg_target_driven}

Blind anonymisation algorithms (e.g., Mondrian partitioning) split the data on quasi-identifiers without considering what the data will be used for afterwards. Target-Driven anonymisation, introduced by Friedman et al.~(\cite{friedman2010kdtree}) and refined by De~Capitani~di~Vimercati et al.~(\cite{de2024dt}), builds a binary decision tree that partitions records by maximising variance reduction on a binary label. Each split selects the QI feature and threshold that best separate the target classes. Each leaf node becomes an equivalence class, and the QI values within it are generalised to their observed $[\min, \max]$ range.

Why does this matter? Records within the same leaf tend to share the same target label, so a classifier trained on the anonymised output can still distinguish positive from negative cases. Blind splitting mixes target classes within leaves and destroys this signal.

VeriX-Anon exists because this distinction is invisible in the output. A data owner who receives an anonymised dataset and a tree structure cannot tell, by inspection alone, whether the cloud used Target-Driven or blind splitting. The verification framework provides that answer.

\subsection{Merkle Trees and Hash-Based Authentication}
\label{sec:bg_merkle}

A Merkle tree is a binary tree in which every node stores a cryptographic hash~(\cite{merkle1989certified}). Leaf nodes hash their own data content. Internal nodes hash the concatenation of their children's hashes. The root hash therefore commits to the entire structure: change any single leaf or internal node, and every hash on the path to the root changes with it.

In VeriX-Anon, the decision tree \emph{is} the Merkle tree. Leaf nodes hash their generalisation bounds, and internal nodes hash their split feature, split value, and children's hashes. If the cloud substitutes a different splitting algorithm or modifies the tree after construction, the root hash will not match the client's independent re-computation.

One distinction from the standard Merkle use case is worth noting. Blockchain systems typically use Merkle trees for single-element inclusion proofs in $O(\log n)$ time: proving that one transaction exists in the tree without revealing the rest. VeriX-Anon performs full tree re-verification in $O(n/k)$ time, because the client needs to verify the entire tree, not just one record's membership. Section~\ref{sec:complexity} discusses this cost in detail.

\subsection{SHAP Values and Feature Attribution}
\label{sec:bg_shap}

VeriX-Anon uses SHAP values not for model interpretation (their usual purpose) but as a data fingerprint. The idea is straightforward: if the anonymisation preserves the predictive relationships in the data, then a model trained on the anonymised output should attribute importance to the same features, in the same proportions, as a model trained on the original data. If those attributions diverge, something changed in the data's structure.

SHAP (SHapley Additive exPlanations), introduced by Lundberg and Lee~(\cite{lundberg2017shap}), assigns each input feature a contribution score for a given prediction. The method is rooted in cooperative game theory: the Shapley value of feature $j$ for input $x$ is its average marginal contribution across all possible subsets of features. For a model $f$, the SHAP values satisfy $\sum_j \phi_j(x) = f(x) - \mathbb{E}[f(X)]$, meaning they fully decompose the gap between a specific prediction and the model's average output. For tree-based models, the TreeExplainer algorithm computes exact SHAP values in polynomial time by exploiting the tree structure directly.

We use SHAP rather than another attribution method for two properties that matter to verification. First, TreeExplainer returns exact Shapley values for tree models, so the fingerprint is deterministic and reproducible: the client and any auditor recover the same values from the same data. LIME and Integrated Gradients depend on sampling or on a chosen reference baseline, so they do not give a stable fingerprint. Second, Shapley values satisfy local accuracy and consistency, so a change in the data's predictive structure appears as a change in the attribution rather than as sampling noise. Tree gain is deterministic as well, but it is a global training statistic that does not compare distributions between two datasets, which is what verification requires.

\subsection{Wasserstein Distance}
\label{sec:bg_wasserstein}

The magnitude of this divergence is measured by the 1-Wasserstein distance~(WD) (also called the Earth Mover's Distance). For two distributions $P$ and $Q$ over $\mathbb{R}$:

\begin{equation}
W_1(P, Q) = \int_{-\infty}^{\infty} |F_P(x) - F_Q(x)|\, dx
\label{eq:wasserstein_def}
\end{equation}

\noindent where $F_P$ and $F_Q$ are the cumulative distribution functions. Informally, $W_1$ measures the minimum cost of reshaping one distribution into the other, where cost is mass moved times distance travelled. Chaudhury et al.~(\cite{chaudhury2024explainable}) applied it to compare SHAP distributions for model explainability; VeriX-Anon adapts the same metric to compare SHAP distributions \emph{before and after anonymisation}. A small $W_1$ indicates that feature importance was preserved. A large $W_1$ indicates that the cloud's processing altered the data's predictive structure, which is evidence of algorithm substitution. We use the 1-Wasserstein distance rather than a difference of means or a KL divergence for three reasons. It is defined even when the two distributions have different supports, whereas KL divergence is not. It is expressed in the same units as the SHAP values, so a threshold on it is interpretable. And it responds to how far probability mass has to move, so it registers a shift that leaves the mean and variance unchanged, which a summary-statistic comparison would miss. The threshold $\varepsilon$ separating acceptable from suspicious divergence is calibrated empirically and discussed in Sections~\ref{sec:wasserstein} and~\ref{sec:epsilon_sensitivity}.
\begin{figure*}[!h]
\centering
\includegraphics[width=\textwidth]{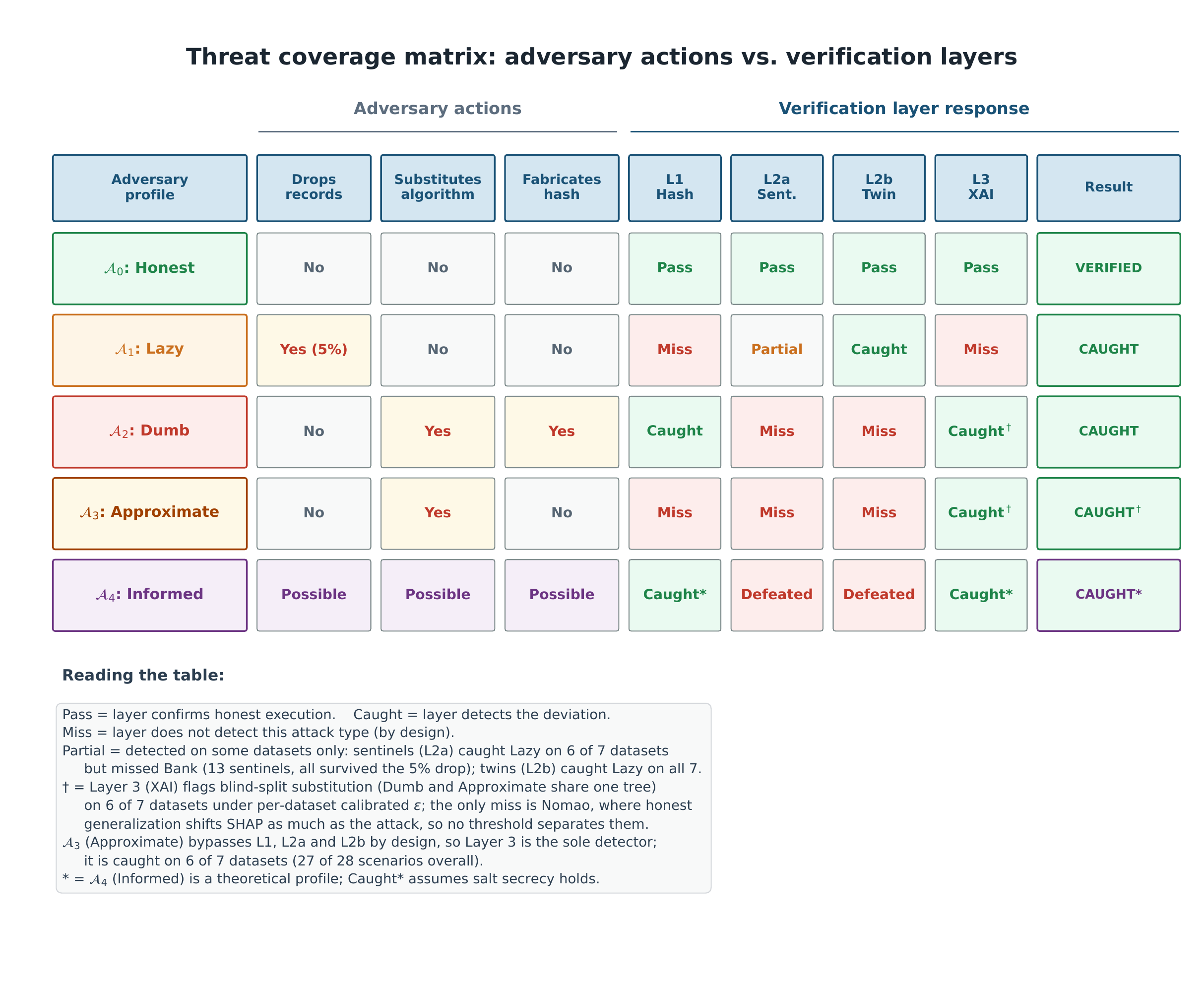}
\caption{Threat coverage matrix. Left columns show each adversary's actions (data dropping, algorithm substitution, hash fabrication). Right columns show which verification layer catches the deviation. No single layer achieves full coverage; the multi-layered design reaches 27 of 28 correct detections across seven datasets under per-dataset threshold calibration. The Approximate adversary ($\mathcal{A}_3$) bypasses Layers~1, 2a, and 2b entirely; Layer~3 (XAI) is the only mechanism that detects it, succeeding on six of seven datasets and failing only on Nomao, where honest generalization shifts SHAP as much as the attack does. $\mathcal{A}_4$ (Informed) is a theoretical profile analysed in Section~\ref{sec:informed_attacker} and is excluded from the detection count; Caught* assumes salt secrecy holds.}
\label{fig:threat_matrix}
\end{figure*}
\section{System Model and Threat Model}
\label{sec:threat_model}

This section defines the three entities in the VeriX-Anon protocol, the adversary profiles against which the framework is evaluated, and the security assumptions that bound the analysis. Figure~\ref{fig:threat_matrix} summarises the adversary action space and per-layer detection coverage.

\subsection{System architecture}
\label{sec:system_arch}

The protocol involves three logical entities. In practice, the client and verification oracle run on the same machine; we separate them for clarity.

\subsubsection{Client (data owner)}

The client holds a dataset $\mathcal{D} = \{(x_i, y_i)\}_{i=1}^{N}$, where $x_i$ is a vector of Quasi-Identifier (QI) attributes and $y_i \in \{0, 1\}$ is a binary target variable. The client wants $k$-anonymity applied to $\mathcal{D}$ via Target-Driven decision tree partitioning, but lacks the computational resources (or prefers not) to perform this operation locally. Before outsourcing, the client injects Boundary Sentinels $\mathcal{S}$ and Twins $\mathcal{T}$ into the dataset, assigns cryptographic TrackerIDs to every record, and transmits the augmented dataset $\mathcal{D}' = \mathcal{D} \cup \mathcal{S} \cup \mathcal{T}$ to the cloud.

\subsubsection{Cloud provider}

The cloud receives $\mathcal{D}'$ and is contractually obligated to:

\begin{enumerate}
\item Build a Target-Driven decision tree that partitions records by maximising variance reduction on the target $y$.
\item Generalise QI values within each leaf to their $[\min, \max]$ range, producing $k$-anonymous equivalence classes.
\item Compute a bottom-up Merkle-style SHA-256 hash over the tree structure.
\item Return three objects to the client: the anonymized dataset $\mathcal{D}^*$, a leaf assignment mapping $\mathcal{L}: \text{TrackerID} \to \text{leaf\_id}$, and the Merkle root hash $H_{\text{root}}$.
\end{enumerate}

The cloud has full access to the feature values and target column of $\mathcal{D}'$. It does not have access to the client's trap manifest $\mathcal{M}$ (the set of sentinel TrackerIDs, the twin pair map, or the client's XAI baseline).

\subsubsection{Verification oracle (client-side)}

After receiving $(\mathcal{D}^*, \mathcal{L}, H_{\text{root}})$ from the cloud, the client executes a four-layer verification audit. This is a local computation: the client never sends verification queries back to the cloud, so the cloud cannot adapt its behaviour based on which checks are being run.

\subsection{Adversary profiles}
\label{sec:adversaries}

We define one honest baseline and three adversary profiles, ordered by increasing deviation from the honest protocol. A fifth profile, the Informed Attacker ($\mathcal{A}_4$), is analysed theoretically in Section~\ref{sec:informed_attacker}.

\subsubsection{Honest cloud ($\mathcal{A}_0$)}

The honest cloud follows the protocol exactly. It processes all records in $\mathcal{D}'$, builds the Target-Driven tree with variance reduction splitting, generalises leaves, computes the correct Merkle hash, and returns the complete result. This profile is the baseline: VeriX-Anon should report \textsc{verified} for all honest executions.

\subsubsection{Lazy cloud ($\mathcal{A}_1$)}

The lazy cloud drops a fraction $\delta = 0.05$ of records from $\mathcal{D}'$ before processing. It then runs the correct Target-Driven algorithm on the reduced dataset $\mathcal{D}'_{\text{red}} \subset \mathcal{D}'$ with $|\mathcal{D}'_{\text{red}}| = (1 - \delta) \cdot |\mathcal{D}'|$. The returned hash is valid for the reduced tree.

Motivation: processing fewer records saves compute. In a pay-per-use cloud model, a 5\% reduction in data volume translates directly to reduced CPU time and memory. The lazy cloud bets that the client will not notice a small number of missing records in a large anonymized output.

Because the tree over the reduced data is internally consistent, the Merkle hash is correct for that tree. Layer~1 (hash verification) will pass. The attack is detectable only through Layer~2 (missing sentinels or twins) or, in some cases, through Layer~3 (SHAP distribution shift if the dropped records alter the feature importance structure).

\subsubsection{Dumb cloud ($\mathcal{A}_2$)}

The dumb cloud replaces Target-Driven splitting with random-feature splitting (Mondrian-style blind partitioning). It processes all records in $\mathcal{D}'$ but ignores the target variable $y$ entirely, splitting on randomly selected features at the median value. After building the blind tree, it fabricates a root hash $H'_{\text{root}} \neq H_{\text{root}}$ to disguise the algorithm substitution.

Motivation: random splitting is cheaper than Target-Driven splitting. Computing variance reduction requires evaluating every candidate feature at every node and reading the target column. Random splitting skips this entirely. The dumb cloud saves computation at the cost of destroying predictive utility in the anonymized output.

Because all records are present and the tree is deterministic (random but consistent), Layer~2 (sentinels and twins) will pass: sentinels are not dropped, and identical twins follow the same random splits into the same leaf. The attack is detectable through Layer~1 (the fabricated hash does not match the client's re-computation) and Layer~3 (SHAP distributions diverge because the blind tree does not preserve the target-feature relationships).

\subsubsection{Approximate cloud ($\mathcal{A}_3$)}
\label{sec:approximate_cloud}

The approximate cloud takes the exact same utility-destroying shortcut as the dumb cloud: it uses random-feature splitting to save compute. But it does not fake the math. It processes every single record and computes a perfectly valid Merkle root hash for its substandard tree.

Motivation: The cloud wants to save CPU cycles but knows the client verifies hashes and row counts. By doing the hashing correctly and keeping all data intact, the cloud effortlessly bypasses Layer~1 (structural checks) and Layer~2 (trap checks). The only way to catch this adversary is to measure the actual machine learning utility of the returned data. This profile exists specifically to test the necessity of Explainable AI fingerprinting. Formally, the Approximate cloud defeats Layers~1 and~2 with certainty, so its success reduces to one condition on Layer~3: it evades detection if and only if the maximum per-feature Wasserstein distance of its blind tree stays below the client's threshold, $\max_f W_f < \varepsilon_d$. Section~\ref{sec:calibration} measures this condition on every dataset. It holds only where honest generalization already perturbs SHAP as much as the attack, which is why the adversary succeeds on exactly one of the seven datasets.

\subsubsection{Adversary limitations and the informed attacker ($\mathcal{A}_4$)}
\label{sec:informed_attacker}

The four profiles above assume the adversary does not know sentinels and twins exist in the outsourced data. A stronger adversary, the Informed Attacker ($\mathcal{A}_4$), knows the client embeds traps and tries to remove them before processing.

Cryptographic TrackerIDs (SHA-256 with a secret salt) make sentinels and twins statistically indistinguishable from genuine records. The TrackerID is a deterministic hash of the salt, the record's role, its index, and its feature vector. Without the salt, the cloud cannot map TrackerIDs to trap records. The IDs themselves are uniformly distributed 256-bit strings. No statistical test on the ID values can separate traps from real data.

Boundary Sentinels do have a structural tell. Because we place them near the decision boundary ($P \in [0.45, 0.55]$), they might cluster in specific regions of the feature space. An attacker with a clean copy of the original public dataset could run outlier detection to flag perturbed records in those dense boundary regions. If the adversary successfully drops the sentinels, Layer~2a becomes useless.

This is exactly why VeriX-Anon uses multiple layers. If the informed attacker defeats Layer~2 completely by removing all sentinels and twins, Layer~1 (Authenticated Decision Tree, ADT Hash) and Layer~3 (XAI Fingerprint) still catch the deviation. Layer~1 flags structural changes to the tree. Layer~3 flags utility degradation. To beat the entire framework, the informed attacker has to identify and remove all traps, reproduce the exact Target-Driven tree structure to satisfy the hash check, and preserve SHAP distributions within $\varepsilon$. If an attacker does all that compute, they have effectively run the honest protocol anyway.

We leave formal analysis of an adaptive informed attacker (one that observes verification outcomes and tweaks its approach across multiple rounds) to future work.
\subsection{Security assumptions}
\label{sec:assumptions}

The verification guarantees of VeriX-Anon rest on four assumptions:

\begin{enumerate}
\item \textbf{Limited client compute.} The client retains enough local resources to train a Random Forest on 10\% of $\mathcal{D}$ (for sentinel generation) and an XGBoost model on 2,000 rows (for XAI fingerprinting). Both operations complete in seconds on commodity hardware. The client does \emph{not} need the resources to build the full anonymization tree, which is the point of outsourcing.

\item \textbf{SHA-256 collision resistance.} The Merkle authentication in Layer~1 relies on the assumption that SHA-256 is collision-resistant. An adversary that can find two distinct tree structures producing the same root hash could defeat Layer~1. Under standard cryptographic assumptions, this requires $O(2^{128})$ work, which is computationally infeasible.

\item \textbf{Trap indistinguishability.} The cloud cannot distinguish sentinel or twin records from genuine records without access to the client's secret salt. TrackerIDs are deterministic SHA-256 hashes that appear uniformly random to any party without the salt. Feature-level indistinguishability depends on the perturbation being small enough that sentinels fall within the natural variation of the dataset (enforced by the $0.05 \cdot \sigma_j$ perturbation bound in Equation~\ref{eq:sentinel_perturb}).

\item \textbf{Secret trap manifest.} The client's trap manifest $\mathcal{M}$ (sentinel IDs, twin pair map, XAI baseline distributions) is never transmitted to the cloud. In particular, the cryptographic salt used in TrackerID generation (Equation~\ref{eq:tracker_id}) remains client-side at all times; if the cloud were to obtain the salt, it could recompute TrackerIDs and identify which records are sentinels, twins, or genuine, defeating Layer~2 entirely. The cloud receives only $\mathcal{D}'$ (the augmented dataset with TrackerIDs and target column). All verification is performed client-side after the cloud returns its output.
\end{enumerate}
\begin{figure*}[!h]
\centering
\includegraphics[width=\textwidth]{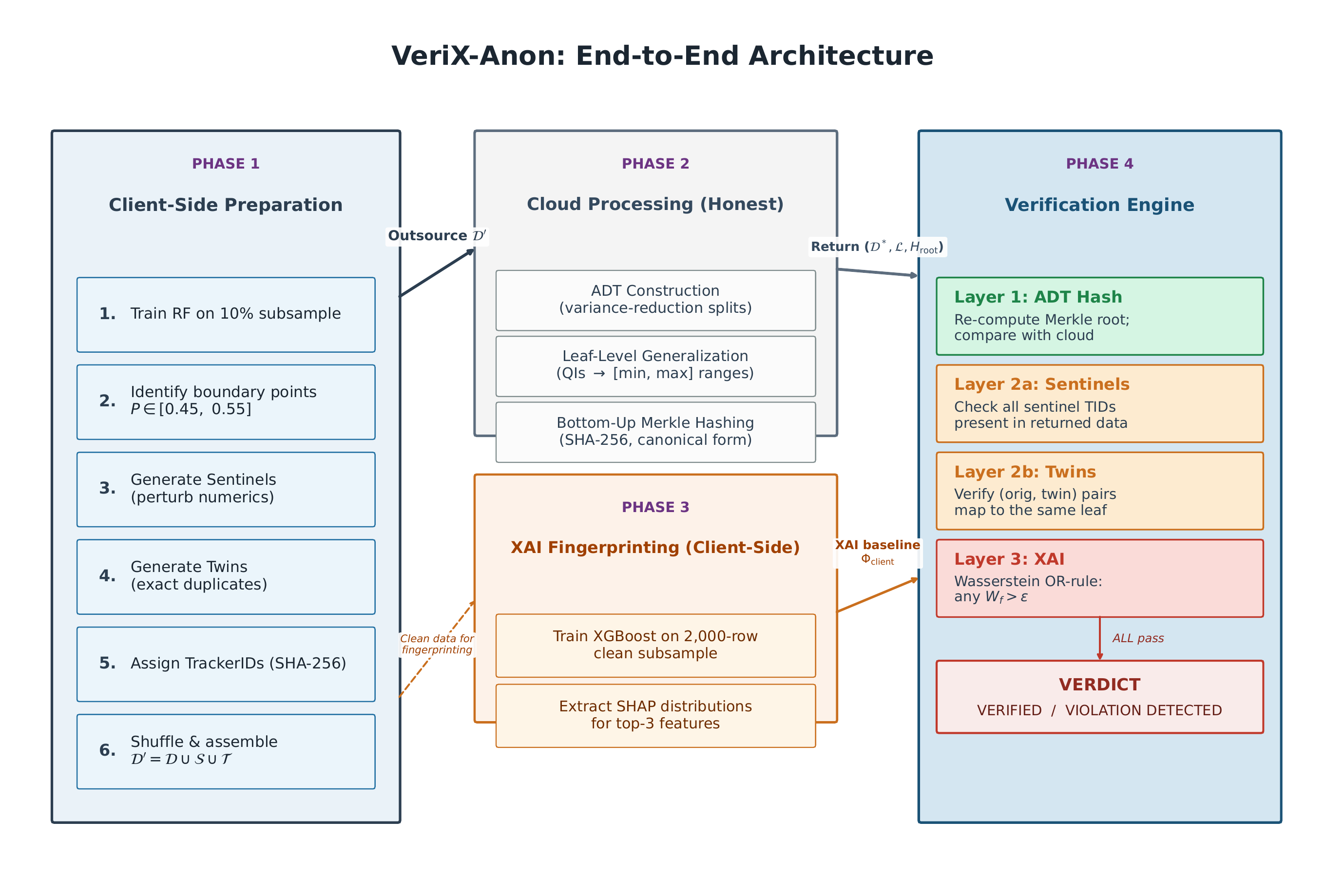}
\caption{VeriX-Anon end-to-end architecture. The client prepares the dataset
with embedded traps (Phase~1), outsources to the cloud for authenticated
Target-Driven anonymization (Phase~2), and runs the four-layer verification
engine (Phase~4) against a locally computed XAI baseline (Phase~3).}
\label{fig:architecture}
\end{figure*}

\section{Proposed Methodology: The VeriX-Anon Framework}
\label{sec:methodology}

VeriX-Anon operates in four sequential phases. The client first prepares the dataset by injecting cryptographically tracked traps (Phase~1). The cloud then performs authenticated Target-Driven anonymization and returns the result with a Merkle root hash (Phase~2). The client independently computes an XAI fingerprint on its local data before outsourcing (Phase~3). Finally, the client runs a four-layer verification engine that cross-checks the cloud's output against all three verification mechanisms (Phase~4). Figure~\ref{fig:architecture} illustrates the end-to-end architecture.

We use the term \emph{mathematically verifiable} to describe a system in which every verification layer has a quantifiable guarantee: Layer~1 provides deterministic correctness (any structural tampering is detected with probability~1 under the collision-resistance assumption of SHA-256), Layer~2 provides probabilistic completeness (sentinel evasion probability is bounded by Equation~\ref{eq:sentinel_evasion} and twin consistency is deterministic), and Layer~3 provides empirically calibrated utility verification (violation detection above the threshold $\varepsilon$ with sensitivity governed by the OR-rule in Equation~\ref{eq:or_rule}). Each layer's detection properties are formally characterised, even where the underlying mechanism is probabilistic or empirical rather than deterministic. We are explicit about the kind of guarantee each layer offers. Only Layer~1 is cryptographic: it detects any structural change with probability~1 under SHA-256 collision resistance. Layer~2 is a probabilistic guarantee with an evasion bound (Equation~\ref{eq:sentinel_evasion}), and Layer~3 is an empirical, calibrated test, not a proof. When we call the framework mathematically verifiable, we mean each layer carries a quantifiable detection property, not that every layer provides a cryptographic proof.

\subsection{Phase 1: Client-Side Preparation}
\label{sec:phase1}

Before outsourcing, the client embeds two types of verifiable structures into the dataset: Boundary Sentinels (probabilistic traps near the decision boundary) and Twins (deterministic duplicate pairs). Both are tagged with cryptographic TrackerIDs that the cloud cannot distinguish from genuine records.

\subsubsection{Boundary Sentinel Generation}
\label{sec:sentinel_gen}

The client trains a Random Forest classifier ($B = 50$ trees, $\text{max\_depth} = 5$) on a 10\% stratified subsample of the original dataset $\mathcal{D}$. For each record $x_i \in \mathcal{D}$, the trained model produces a class-1 probability $P(y = 1 \mid x_i)$. Records falling within the uncertainty band $P \in [0.45, 0.55]$ sit near the decision boundary, where they are most sensitive to changes in the splitting logic.

From this boundary set, the client selects up to $\lceil 0.02 \cdot N \rceil$ candidates and perturbs only their numerical columns:

\begin{equation}
x'_j = \text{clip}\!\left(x_j + \mathcal{N}(0,\; 0.05 \cdot \sigma_j),\; x_j^{\min},\; x_j^{\max}\right)
\label{eq:sentinel_perturb}
\end{equation}

\noindent where $\sigma_j$ is the population standard deviation of column $j$, and the clip operation ensures that values fall within the range of that column. Categorical columns are not modified. Integer-type columns are rounded at this stage. This generates synthetic data that is statistically valid but placed in a position where any variation from Target-Driven logic, such as random splitting, will move them to incorrect leaves.

The number of sentinels depends on the class distribution within the dataset. Balanced datasets will have more sentinels than imbalanced ones. In our experiments, Adult Income had 501 boundary candidates, 160 within the 2\% limit, while Bank Marketing had only 13. The implications are given in Section~\ref{sec:sentinel_density}.

\subsubsection{Twin Generation}
\label{sec:twin_gen}

The client selects 5\% of genuine records uniformly at random and creates exact duplicates. Under a deterministic anonymization algorithm, two identical feature vectors must traverse the same tree path and land in the same leaf. If the cloud's returned leaf assignments place any (original, twin) pair in different leaves, this constitutes proof that the cloud used a non-deterministic or randomized method.

Formally, let $x_i$ be a genuine record and $x_i' = x_i$ be its twin. Under deterministic tree $\mathcal{T}$:

\begin{equation}
\mathcal{T}(x_i) = \mathcal{T}(x_i') \quad \forall\; (x_i, x_i') \in \text{TwinPairs}
\label{eq:twin_consistency}
\end{equation}

\noindent Any violation of Equation~\ref{eq:twin_consistency} is a sufficient condition for detecting non-deterministic processing.

A potential concern is that exact duplicates could be detected by an adversary running standard deduplication on the outsourced data. In datasets with continuous numerical features, exact duplicates are statistically rare in natural data, and their presence could raise suspicion. Two mitigations exist. First, the twin injection rate (5\%) is low enough that duplicates are plausible as natural repetition in large administrative datasets (e.g., repeated hospital visits or duplicate survey entries). Second, future versions could replace exact twins with near-duplicates, applying a small perturbation (e.g., $\pm 1$ in the least significant digit of numerical features) that preserves the same-leaf guarantee under deterministic tree splitting with quantised thresholds. Formal analysis of near-duplicate twins is left to future work.
\subsubsection{Cryptographic Tracking IDs}
\label{sec:tracker_ids}

Each record (genuine, sentinel, or twin) receives a deterministic SHA-256 identifier:

\begin{equation}
\text{TID}_i = \text{SHA-256}\!\left(\texttt{salt} \;\|\; \texttt{role}_i \;\|\; i \;\|\; \text{bytes}(x_i)\right)
\label{eq:tracker_id}
\end{equation}

\noindent where $\texttt{salt}$ is a secret string known only to the client, $\texttt{role}_i \in \{\texttt{genuine}, \texttt{sentinel}, \texttt{twin}\}$, and $\text{bytes}(x_i)$ is the byte-level representation of the record's feature vector. Because the hash is deterministic, the client can always recompute any TrackerID without storing random nonces.

For twin pairs, the client maintains a local mapping $\texttt{twin\_pair\_map}[\text{TID}_{\text{orig}}] = \text{TID}_{\text{twin}}$ that links each original record to its duplicate. This mapping, along with the set of sentinel TrackerIDs, forms the client's \emph{trap manifest} $\mathcal{M}$, which is never shared with the cloud.

\subsubsection{Dataset Assembly and Outsourcing}
\label{sec:assembly}

The client concatenates genuine records, sentinels, and twins into a single outsourced dataset:

\begin{equation}
\mathcal{D}' = \mathcal{D} \cup \mathcal{S} \cup \mathcal{T}
\label{eq:dataset_assembly}
\end{equation}

\noindent The combined dataset is shuffled using a seeded pseudorandom permutation before transmission. Shuffling prevents the cloud from inferring trap positions based on row ordering. Each row carries its TrackerID and the binary target column; the cloud has no information about which rows are genuine, sentinels, or twins.

\subsection{Phase 2: Authenticated Target-Driven Anonymization (ADT)}
\label{sec:phase2}

The cloud receives $\mathcal{D}'$ and is contracted to perform Target-Driven $k$-anonymization. This section describes the honest protocol; adversarial deviations are defined in Section~\ref{sec:threat_model}.

\subsubsection{Target-Driven Decision Tree Construction}
\label{sec:tree_construction}
The cloud builds a binary decision tree that partitions records to maximize within-leaf target homogeneity. At each internal node, the cloud selects the feature $f^*$ and split value $s^*$ (the column median) that maximise variance reduction:\footnote{In the reference implementation, the parent node variance is computed using the pandas \texttt{Series.var()} method (sample variance, $\mathrm{ddof}=1$), while child node variances are computed using NumPy array \texttt{.var()} (population variance, $\mathrm{ddof}=0$). This inconsistency has negligible effect on split selection because all candidate splits at a given node share the same parent variance as the reference value, and child nodes contain at least $2k = 10$ records, where the difference between $n$ and $n{-}1$ denominators is under 10\%. Standardising to a single convention would not change any split decision in the experiments reported here.}

\begin{equation}
\Delta\sigma^2 = \sigma^2_{\text{parent}} - \frac{n_L \cdot \sigma^2_L + n_R \cdot \sigma^2_R}{n_{\text{parent}}}
\label{eq:variance_reduction}
\end{equation}

\noindent where $\sigma^2_{\text{parent}}$, $\sigma^2_L$, and $\sigma^2_R$ are the target variances of the parent, left child, and right child respectively, and $n_L$, $n_R$, $n_{\text{parent}}$ are the corresponding record counts.

Tree construction terminates when any of the following conditions holds:

\begin{enumerate}
\item The number of records within the node is less than $2 \times \text{min\_leaf}$, where $\text{min\_leaf} = 2k$ (i.e., the node contains fewer than $4k$ records). This ensures that any binary split produces children with at least $2k$ records each, satisfying the $k$-anonymity guarantee.
\item The node is pure, i.e., $\sigma^2 = 0$, which means that all records within this node have the same target value.
\item The depth of the tree exceeds a safety limit of 50. This limit prevents unbounded recursion in pathological cases (e.g., features with many unique values producing long chains of binary splits). In practice, the deepest tree observed in our experiments had 11 levels (Diabetes, 8,000 rows, $k = 5$). The limit of 50 is never reached during normal operation and serves only as a safeguard.
\end{enumerate}

The choice of using the median ensures that we do not have any "degenerate" splits that could result in a leaf with too few records to satisfy the $k$-anonymity condition.

\subsubsection{Leaf-Level Generalization}
\label{sec:generalization}

Each leaf node generalises its Quasi-Identifier (QI) columns by replacing individual values with the observed $[\min, \max]$ range within that leaf. All records assigned to the same leaf become indistinguishable on every QI attribute. Because each leaf contains at least $2k$ records (enforced by the stopping rule), the resulting partition satisfies $k$-anonymity.

For downstream compatibility (e.g., training a classifier on anonymized data), each generalised value can be reconstructed as the range midpoint:

\begin{equation}
x_{\text{mid}}^{(j)} = \frac{x_{\min}^{(j)} + x_{\max}^{(j)}}{2}
\label{eq:midpoint}
\end{equation}

Here $x_{\min}^{(j)}$ and $x_{\max}^{(j)}$ are the smallest and largest values of feature $j$ within the leaf, so the midpoint is the single representative value a downstream model sees in place of the generalized range.

\subsubsection{Merkle-Style Tree Authentication}
\label{sec:merkle_auth}

After building the tree and generalising leaves, the cloud computes a bottom-up SHA-256 hash over the entire tree structure. This produces a single root hash $H_{\text{root}}$ that cryptographically commits to every split decision and every leaf's generalisation bounds.

For leaf nodes:
\begin{equation}
H_{\ell} = \text{SHA-256}\!\left(\texttt{"LEAF|"} \;\|\; |\ell| \;\|\; \texttt{"|"} \;\|\; \text{canonical\_bounds}(\ell)\right)
\label{eq:leaf_hash}
\end{equation}

\noindent where $|\ell|$ is the number of records in leaf $\ell$, and $\text{canonical\_bounds}(\ell)$ is a deterministic string encoding of the leaf's generalisation ranges, sorted alphabetically by feature name with values rounded to 6 decimal places and pipe-separated.

\begin{equation}
\begin{aligned}
H_v = \text{SHA-256}(&\texttt{"INTERNAL|"} \;\|\; f_v \;\|\; \texttt{"|"} \\
                    &\;\|\; s_v \;\|\; \texttt{"|"} \;\|\; H_L \\
                    &\;\|\; \texttt{"|"} \;\|\; H_R)
\end{aligned}
\label{eq:internal_hash}
\end{equation}

\noindent where $f_v$ is the split feature, $s_v$ is the split value (rounded to 6 decimal places), and $H_L$, $H_R$ are the hashes of the left and right children. The canonical formatting (fixed decimal precision, alphabetical sorting) eliminates platform-dependent floating-point representation issues.

The cloud returns three objects to the client: (1) the anonymized dataset $\mathcal{D}^*$, (2) a leaf assignment mapping $\mathcal{L}: \text{TrackerID} \to \text{leaf\_id}$, and (3) the root hash $H_{\text{root}}$.

\subsection{Phase 3: XAI Fingerprinting}
\label{sec:phase3}

Before outsourcing, the client computes an expected ``fingerprint'' of its data's predictive structure. After receiving the cloud's output, the client computes a second fingerprint on the anonymized data and measures the divergence. Large divergence indicates that the cloud's processing destroyed the predictive logic encoded in the original data.

\subsubsection{Client Baseline Computation}
\label{sec:client_baseline}

The client trains an XGBoost classifier (100 estimators, max depth 6, learning rate 0.1) on a 2,000-row subsample of the clean data $\mathcal{D}$. Using TreeExplainer~(\cite{lundberg2017shap}), the client extracts SHAP value distributions for the top-3 features ranked by mean absolute SHAP value ($\text{mean}|\phi_f|$). These distributions form the client's baseline fingerprint $\Phi_{\text{client}} = \{\phi_{\text{client}}^{(f)}\}_{f \in \text{top-3}}$.

The 2,000-row subsample keeps SHAP computation tractable regardless of dataset size and makes the XAI layer $O(1)$ with respect to $n$. The choice of 2,000 rows balances two competing concerns. A smaller subsample (e.g., 500 rows) risks unstable SHAP estimates, particularly for features with heavy-tailed distributions, because TreeExplainer's output variance scales inversely with sample size. A larger subsample (e.g., 5,000 rows) increases computation time without proportionally improving fingerprint stability, since XGBoost with 100 estimators converges in its feature importance rankings well below 2,000 rows for the datasets tested. In our experiments, the top-3 feature rankings were consistent across 1,000-row and 2,000-row subsamples for both Adult Income and Bank Marketing. The 5,000-row Experiment~3 subsample produced a slightly different top-3 ranking (\texttt{age}, \texttt{marital-status}, \texttt{relationship} vs.\ \texttt{age}, \texttt{relationship}, \texttt{education-num}), which we attribute to subsample composition rather than instability. A formal sensitivity analysis across subsample sizes is a valuable direction for future work.

\subsubsection{Cloud Output Evaluation}
\label{sec:cloud_eval}

Upon receiving the anonymized dataset $\mathcal{D}^*$, the client flattens generalised ranges to midpoint values using Equation~\ref{eq:midpoint}, then trains an identical XGBoost model on a 2,000-row subsample and extracts SHAP distributions for the same top-3 features: $\Phi_{\text{cloud}} = \{\phi_{\text{cloud}}^{(f)}\}_{f \in \text{top-3}}$.

\subsubsection{Wasserstein Distance Comparison and Violation Rule}
\label{sec:wasserstein}

For each top-3 feature $f$, the client computes the 1-Wasserstein distance (Earth Mover's Distance) between the client and cloud SHAP distributions:

\begin{equation}
W_f = W_1\!\left(\phi_{\text{client}}^{(f)},\; \phi_{\text{cloud}}^{(f)}\right)
\label{eq:wasserstein}
\end{equation}

The client applies an OR-rule: if \emph{any} single feature's Wasserstein distance exceeds the threshold $\varepsilon$, the cloud is flagged for an algorithmic integrity violation:

\begin{equation}
\text{XAI\_Violation} = \exists\; f \in \text{top-3} \;\text{s.t.}\; W_f > \varepsilon
\label{eq:or_rule}
\end{equation}

The OR-rule is strictly more sensitive than an averaging rule. If the cloud substitutes a utility-destroying algorithm (e.g., random splitting), it may damage one feature's SHAP distribution severely while leaving others relatively intact. Averaging could mask this single-feature damage; the OR-rule catches it.

The value of $\varepsilon$ is set to 0.45 after being empirically tuned for the default operating point $k = 5$. At this $k$-value, the maximum honest per-feature Wasserstein distance is 0.4436 (\texttt{relationship}), sitting 0.0064 below $\varepsilon$. The minimum adversarial per-feature distance that triggers the OR-rule in the Experiment~3 feature set is 0.4574 (\texttt{relationship} under the Blind adversary at $k = 5$),giving a separation margin of 0.0138. We emphasise that $\varepsilon$ must be calibrated per contracted $k$-value: at $k = 10$, 12, and 15, the honest per-feature WD for \texttt{marital-status} exceeds 0.45 (reaching 0.5744 at $k = 12$), which would produce false positives under the current global threshold. In a production deployment, the client would calibrate $\varepsilon$ for the specific $k$-value contracted with the cloud, not across the full $k$-sweep.

We emphasise that $\varepsilon$ is not a learned model parameter and its calibration does not constitute data leakage in the machine learning sense. In a real deployment, the client would establish $\varepsilon$ during a one-time calibration phase: the client runs the Target-Driven anonymisation locally on a small representative sample (e.g., 1{,}000 rows) at the contracted $k$-value, computes the honest SHAP Wasserstein distances, and sets $\varepsilon$ above the observed honest maximum with a safety margin. This calibration requires the client to anonymise only a small sample locally (seconds of compute), not the full dataset. The adversarial distribution is not needed for calibration because any $\varepsilon$ set above the honest baseline will flag deviations that exceed it. The $k$-sweep in Section~\ref{sec:utility} serves to validate this approach across a range of $k$-values; it is not used to fit $\varepsilon$ to adversarial outputs. Section~\ref{sec:calibration} formalises this as a per-dataset rule (Equation~\ref{eq:eps_cal}) and reports its effect on detection.
\subsection{Phase 4: Multi-Layered Client Verification Engine}
\label{sec:phase4}

Upon receipt of the cloud's output $(\mathcal{D}^*, \mathcal{L}, H_{\text{root}})$, the client performs four independent verification procedures. The motivation behind this is the "Swiss Cheese" model, where each layer protects against a different type of deviation, with the combination protecting against deviations not covered by any single layer.

\subsubsection{Layer 1: ADT Hash Verification}
\label{sec:layer1}

The client reconstructs the Merkle hash tree from the returned anonymized data and tree structure using Equations~\ref{eq:leaf_hash} and~\ref{eq:internal_hash}. If the hash value at the root of the reconstructed tree matches the reported hash value from the cloud $H_{\text{root}}$, it is certain that the tree structure is unchanged. Any difference between the two hash values indicates that the cloud either altered the tree structure or fabricated a hash value.

This layer is deterministic; that is, there are no false positives or false negatives for structural tampering. The complexity of this layer is $O(n/k)$.

\subsubsection{Layer 2a: Sentinel Presence Check}
\label{sec:layer2a}

At this layer, the client checks if the sentinels from $\mathcal{M}$ are present in $\mathcal{D}^*$. The absence of sentinels indicates that the cloud has discarded data during processing.

\begin{equation}
\text{Sentinel\_Pass} = \left(\forall\; \text{TID}_s \in \mathcal{S}_{\text{IDs}}: \text{TID}_s \in \mathcal{D}^*.\text{TrackerIDs}\right)
\label{eq:sentinel_check}
\end{equation}

The probability that a data-dropping adversary (dropping fraction $\delta$) evades all $|S|$ sentinels is:

\begin{equation}
P_{\text{evade}} = \left(1 - \frac{|S|}{|\mathcal{D}'|}\right)^{|\mathcal{D}'| \cdot \delta}
\label{eq:sentinel_evasion}
\end{equation}

When $|S|$ is large compared to $|\mathcal{D}'|$, evasion probability falls quickly. Yet, when class imbalance results in a small number of boundary candidates (e.g., in the Bank Marketing problem, $|S|=13$ compared to $|\mathcal{D}'|=8,413$), evasion probability can be substantial (Section~\ref{sec:sentinel_density}).
Equation~\ref{eq:sentinel_evasion} assumes uniform random dropping. If the adversary uses a non-uniform strategy (e.g., preferentially dropping records from the majority class to reduce compute while minimally affecting the tree), the evasion probability would differ. Majority-class-biased dropping would be less likely to hit sentinels that cluster near the decision boundary (which is class-balanced by construction), potentially reducing evasion probability relative to the uniform case.

Complexity: $O(|S|)$, where $|S| \leq 0.02n$.

\subsubsection{Layer 2b: Twin Leaf Consistency Check}
\label{sec:layer2b}

For every (original, twin) pair in $\texttt{twin\_pair\_map}$, the client verifies that both TrackerIDs map to the same leaf in $\mathcal{L}$:

\begin{equation}
\begin{aligned}
\text{Twin\_Pass} = \Big(&\forall\; (\text{TID}_o, \text{TID}_t) \in \texttt{twin\_pair\_map}: \\
                        &\mathcal{L}[\text{TID}_o] = \mathcal{L}[\text{TID}_t]\Big)
\end{aligned}
\label{eq:twin_check}
\end{equation}

This check is independent of Layer 1. A Lazy adversary that drops 5\% of records but runs the correct algorithm on the remainder will produce a valid hash for the reduced tree, passing Layer~1. However, dropped twins cause missing entries in $\mathcal{L}$, and any non-deterministic processing (even with all records present) causes leaf mismatches.

Complexity: $O(|T|)$, where $|T| = 0.05n$.

\subsubsection{Layer 3: XAI Fingerprint Verification}
\label{sec:layer3}

The client evaluates the cloud's anonymized output using the procedure described in Section~\ref{sec:wasserstein}. If any top-3 feature's Wasserstein distance exceeds $\varepsilon$, the layer flags a violation.

This layer is completely orthogonal to Layers~1 and~2. A Dumb adversary fakes the root hash and gets caught by Layer~1. An Approximate adversary, however, computes a mathematically perfect hash for a utility-destroying tree. Because the hash is valid and no data is dropped, Layers~1 and~2 pass the output. Layer~3 is the only mechanism that catches this utility degradation. Conversely, a Lazy adversary that drops just 5\% of records might produce Wasserstein distances within $\varepsilon$, slipping past Layer~3 but getting caught by Layer~2.

Complexity: $O(1)$ relative to $n$, since SHAP computation uses a fixed 2,000-row subsample.

\subsubsection{Verdict Aggregation}
\label{sec:verdict}

The final verdict is the conjunction of all four checks:

\begin{equation}
\text{Verdict} = \bigwedge_{i=1}^{4} \text{Layer}_i\!\left(\mathcal{D}^*, \mathcal{L}, H_{\text{root}}, \mathcal{M}\right)
\label{eq:verdict}
\end{equation}

\noindent where $\mathcal{M}$ is the client's trap manifest (sentinel IDs, twin pair map, and XAI baseline). The system reports \textsc{verified} if and only if all layers pass. A failure in any single layer produces \textsc{violation detected}, and the audit log records which specific layer(s) triggered.

\subsection{Complexity Analysis}
\label{sec:complexity}

Table~\ref{tab:complexity} summarises the per-layer verification cost. The tree hash verification visits every node in the decision tree. Because each leaf contains at least $2k$ records, the number of leaves is at most $n / (2k)$, and the total number of nodes (leaves plus internals) is $O(n/k)$. This is not $O(\log n)$: a standard single-element Merkle proof traverses a root-to-leaf path of length $O(\log n)$, but VeriX-Anon performs \emph{full} re-verification of the entire tree, which requires visiting every node. The distinction matters because $n/k$ can be substantially larger than $\log n$ for large datasets with small $k$.

The sentinel check and twin check are linear with respect to the number of injected traps, which is a constant proportion of $n$. The XAI layer trains a model on 2,000 rows exactly and uses Wasserstein distance over three feature distributions, which is constant with respect to $n$.
\begin{table}[!h]
\centering
\caption{Client-side verification complexity by layer. Total cost is dominated by the $O(n/k)$ hash traversal. The XAI layer is constant with respect to $n$ because SHAP computation uses a fixed subsample of 2,000 rows.}
\label{tab:complexity}

\resizebox{\columnwidth}{!}{
\begin{tabular}{|l|l|l|}
\hline
\textbf{Layer} & \textbf{Operation} & \textbf{Complexity} \\
\hline
Layer 1 (ADT Hash) & Full tree hash re-computation & $O(n/k)$ \\
\hline
Layer 2a (Sentinels) & TrackerID presence lookup & $O(|S|) \approx O(0.02n)$ \\
\hline
Layer 2b (Twins) & Leaf assignment comparison & $O(|T|) \approx O(0.05n)$ \\
\hline
Layer 3 (XAI) & SHAP extraction + Wasserstein & $O(1)$ \\
\hline
\textbf{Total} & & $O(n/k)$ \\
\hline
\end{tabular}
}
\end{table}

In reality, for smaller values of $n$, the XAI cost is dominant at around 0.5s to train the model and compute SHAP. Conversely, for larger values of $n$, the hash traversal cost is dominant. Section~\ref{sec:scalability} shows that verification time remains sub-second for $n = 10^6$.

\section{Experimental Results and Evaluation}
\label{sec:experiments}

This section reports three experiments. Experiment~1 evaluates detection accuracy across seven datasets and 4~cloud profiles (7~honest + 21~adversarial = 28 scenarios total). Experiment~2 measures client-side verification time from $n = 10{,}000$ to $n = 1{,}000{,}000$. Experiment~3 quantifies the utility-privacy trade-off using an 11-point $k$-sweep on all seven datasets with paired statistical tests. All code runs as a single reproducible Kaggle notebook, available from the corresponding author upon reasonable request.

\subsection{Experimental Setup}
\label{sec:setup}

\subsubsection{Datasets}

We evaluate VeriX-Anon on seven publicly available datasets chosen to span distinct domains and, deliberately, distinct class-balance and signal regimes. The lineup was selected on two properties fixed before any detection run, namely class balance and the AUC of a reference classifier, so it maps the operating envelope of the framework rather than a favourable subset. It spans a high-dimensional case (Nomao, 118 features), two financial-transaction datasets, an energy dataset, and a physics dataset. Table~\ref{tab:datasets} lists them.

\begin{table}[!h]
\centering
\caption{The seven evaluation datasets. Source gives the OpenML identifier or repository. Neg/Pos is the percentage class balance of the binary target. QIs is the number of quasi-identifier columns.}
\label{tab:datasets}
\resizebox{\columnwidth}{!}{
\begin{tabular}{|l|l|l|r|r|c|}
\hline
\textbf{Dataset} & \textbf{Domain} & \textbf{Source} & \textbf{Rows} & \textbf{QIs} & \textbf{Neg/Pos} \\
\hline
Adult Income & Societal & OpenML 1590 & 48,842 & 14 & 76/24 \\
\hline
Bank Marketing & Financial & OpenML 1461 & 45,211 & 16 & 88/12 \\
\hline
Diabetes 130-US & Medical & UCI repo & 101,766 & 16 & 89/11 \\
\hline
Electricity & Energy & OpenML 151 & 45,312 & 8 & 58/42 \\
\hline
Nomao & Web (high-dim) & OpenML 1486 & 34,465 & 118 & 71/29 \\
\hline
Credit Default & Financial & OpenML 42477 & 30,000 & 23 & 78/22 \\
\hline
MagicTelescope & Physics & OpenML 1120 & 19,020 & 10 & 65/35 \\
\hline
\end{tabular}
}
\end{table}

The binary targets are, respectively: annual income above \$50K; term-deposit subscription; hospital readmission within 30 days; a rise in the electricity price; whether two records describe the same place; default on the next payment; and gamma-ray versus hadron signal. For Bank Marketing~(\cite{moro2011bank}), OpenML encodes column names as V1 through V16; we use the original UCI feature names (age, job, marital, education, default, balance, housing, loan, contact, day, month, duration, campaign, pdays, previous, poutcome) throughout this paper. The full mapping is provided in Appendix~B. For Diabetes 130-US, missing values encoded as ? in the original CSV were imputed using column-wise mode for categorical features and median for numerical features.

\subsubsection{Configuration}

All experiments use the parameters in Table~\ref{tab:config}. For Experiment~1, each dataset is subsampled to 8,000 rows. This size is chosen to balance two constraints: (a) the 28-scenario evaluation (7 datasets $\times$ 4 cloud profiles) requires 28 full anonymisation runs plus 28 verification audits, each involving tree construction, SHAP extraction, and hash computation; and (b) the verification mechanisms (hash checking, sentinel presence, twin consistency, SHAP comparison) operate identically regardless of dataset size, since they depend on structural properties (tree topology, record presence, feature distributions) rather than raw row count. 

The scalability experiment (Section~\ref{sec:scalability}) separately confirms that verification time remains sub-second at $n = 10^6$. Experiment~3 uses 5,000 rows for the same reason: the 11-point $k$-sweep requires 22 anonymisation runs (11 $\times$ 2 methods), and larger subsamples would not change the relative F1 or WD comparisons between Target-Driven and Blind anonymisation. For Experiment~3, the Adult Income dataset is subsampled to 5,000 rows to allow the 11-point $k$-sweep to complete within reasonable time.
\begin{table}[!h]
\centering
\caption{Default experimental configuration. All parameters are fixed across datasets unless stated otherwise.}
\label{tab:config}
\resizebox{\columnwidth}{!}{
\begin{tabular}{|l|l|}
\hline
\textbf{Parameter} & \textbf{Value} \\
\hline
$k$-anonymity parameter & 5 \\
\hline
Wasserstein threshold $\varepsilon$ & 0.45 \\
\hline
SHAP subsample size & 2,000 rows \\
\hline
Sentinel injection ratio & 2\% of $N$ \\
\hline
Twin injection ratio & 5\% of $N$ \\
\hline
Random Forest (sentinel gen.) & 50 trees, max depth 5 \\
\hline
XGBoost (XAI fingerprint) & 100 estimators, max depth 6 \\
\hline
ADT max tree depth & 50 \\
\hline
Bootstrap resamples & 10,000 \\
\hline
Lazy adversary drop fraction $\delta$ & 0.05 \\
\hline
\end{tabular}
}
\end{table}

\subsubsection{Trap Injection Summary}

The trap injection counts for each dataset at $k=5$ are given in Table~\ref{tab:trap_injection}. The number of sentinel yield depends on the class balance. For Adult Income, there are 501 boundary candidates, limited by the 160 sentinels imposed by the 2\% limit. Bank Marketing (88/12) has only 13 boundary points and Diabetes (89/11) only 4, all of which become sentinels; the other four datasets each reach the 160-sentinel cap. This affects the sentinel-based detection power, as discussed in Section~\ref{sec:sentinel_density}.

\begin{table}[!h]
\setlength{\tabcolsep}{4pt}
\renewcommand{\arraystretch}{1.0}
\centering
\caption{Trap injection summary for each dataset at $k = 5$. Boundary points are records with RF prediction probability $P \in [0.45, 0.55]$. Trap ratio is the fraction of the outsourced dataset that consists of sentinels and twins.}
\label{tab:trap_injection}
\resizebox{\columnwidth}{!}{
\begin{tabular}{|l|r|r|r|r|c|}
\hline
\textbf{Dataset} & \textbf{Boundary Pts.} & \textbf{Sentinels} & \textbf{Twins} & \textbf{Outsourced} & \textbf{Trap Ratio} \\
\hline
Adult Income & 501 & 160 & 400 & 8,560 & 6.5\% \\
\hline
Bank Marketing & 13 & 13 & 400 & 8,413 & 4.9\% \\
\hline
Diabetes 130-US & 4 & 4 & 400 & 8,404 & 4.8\% \\
\hline
Electricity & 434 & 160 & 400 & 8,560 & 6.5\% \\
\hline
Nomao & 214 & 160 & 400 & 8,560 & 6.5\% \\
\hline
Credit Default & 381 & 160 & 400 & 8,560 & 6.5\% \\
\hline
MagicTelescope & 567 & 160 & 400 & 8,560 & 6.5\% \\
\hline
\end{tabular}
}
\end{table}

\subsubsection{Software Environment}
All experiments were run on Kaggle's free-tier cloud notebook environment with a single Intel Xeon CPU at 2.20\,GHz, without any GPU acceleration. The software stack we used is listed in Table~\ref{tab:software}. Our framework relies only on standard Python libraries, including \texttt{hashlib} for the SHA-256 hashing algorithm, and widely used open-source tools. We do not rely on any proprietary or custom-compiled software or GPU-specific tools.

\begin{table}[htbp]
\centering
\caption{Software environment for all experiments.}
\label{tab:software}
\resizebox{\columnwidth}{!}{
\begin{tabular}{ll}
\hline
\textbf{Component} & \textbf{Version} \\
\hline
Python & 3.12.12 \\
NumPy & 2.0.2 \\
pandas & 2.3.3 \\
scikit-learn & 1.6.1 \\
XGBoost & 3.2.0 \\
SHAP & 0.50.0 \\
SciPy & 1.16.3 \\
Matplotlib & 3.10.0 \\
\texttt{hashlib} & Python stdlib \\
\hline
Platform & Linux (Kaggle), Intel Xeon @ 2.20\,GHz, CPU-only \\
\hline
\end{tabular}
}
\end{table}

\subsection{Multi-Layered Threat Detection Results}
\label{sec:detection}

\begin{figure*}[!h]
\centering
\includegraphics[width=\textwidth]{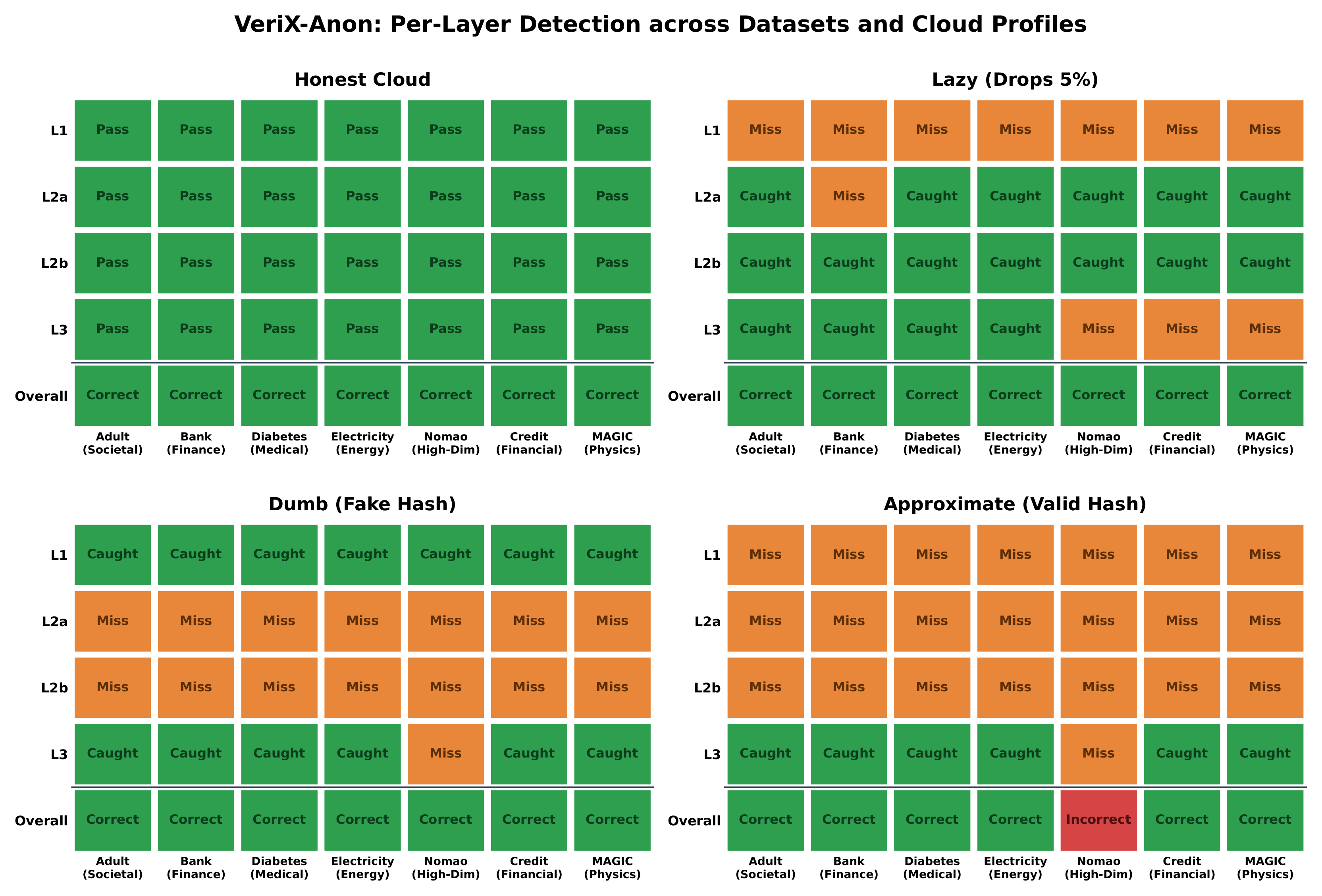}
\caption{Per-layer verification coverage across seven datasets and four cloud profiles (28 scenarios). Green cells mark correct behaviour (pass for honest, caught for malicious); orange cells mark a missed attack on that layer. Layer~3 (XAI) is the only layer that catches the Approximate adversary, and Nomao is the one dataset where it cannot, because there honest generalization perturbs SHAP as much as the attack. Under the per-dataset calibrated threshold the combined framework is correct in 27 of 28 scenarios.}
\label{fig:heatmap}
\end{figure*}

Across the 28 scenarios, VeriX-Anon is correct in 25 under the fixed global threshold $\varepsilon = 0.45$ and in 27 once the threshold is calibrated per dataset (Section~\ref{sec:calibration}). Table~\ref{tab:detection_perf} reports detection performance under both thresholds, and Table~\ref{tab:detection_full} in Appendix~\ref{app:detection} gives the per-scenario, per-layer verdicts. No single layer is correct on more than 23 of the 28 scenarios by itself; only the combination reaches 27. Figure~\ref{fig:heatmap} maps the coverage visually.

\begin{table}[!h]
\centering
\caption{Detection performance over 28 scenarios (7 datasets $\times$ 4 cloud profiles) under the fixed global threshold and the per-dataset calibrated threshold. Precision, recall, and specificity treat a detected deviation as a positive.}
\label{tab:detection_perf}
\resizebox{\columnwidth}{!}{
\begin{tabular}{|l|c|c|c|c|c|c|}
\hline
\textbf{Threshold} & \textbf{Correct} & \textbf{Precision} & \textbf{Recall} & \textbf{Specificity} & \textbf{FP} & \textbf{FN} \\
\hline
Global ($\varepsilon = 0.45$) & 25/28 & 0.95 & 0.90 & 0.86 & 1 & 2 \\
\hline
Calibrated (per dataset) & 27/28 & 1.00 & 0.95 & 1.00 & 0 & 1 \\
\hline
\end{tabular}
}
\end{table}

\subsubsection{Dumb Cloud analysis}
The Dumb adversary substitutes Target-Driven splitting with random-feature Mondrian partitioning and fabricates a root hash. Layer~1 catches it on all seven datasets, because the fabricated hash never matches the client's re-computation. Layer~3 independently flags it on six of seven datasets under the calibrated threshold, since the blind tree pushes at least one top-3 SHAP feature past that dataset's $\varepsilon$ (Table~\ref{tab:per_feature_wd}). The exception is Nomao, whose honest SHAP distribution is already so dispersed that the blind attack does not move it any further (Section~\ref{sec:diabetes_anomaly}). Because Layer~1 is deterministic, the Dumb cloud is caught on every dataset regardless of what Layer~3 does.

Layer 2 (both sentinels and twins) does not detect the Dumb adversary because the adversary processes all records (no dropping of data) and because, although Mondrian partitioning is utility-destroying, it is still deterministic: identical twin vectors have the same random splits to the same leaf.

\subsubsection{Lazy Cloud analysis}
The Lazy adversary drops 5\% of records but runs the correct Target-Driven algorithm on the rest. The reduced tree is internally consistent, so its hash is valid and Layer~1 passes on every dataset. Layer~3 is unreliable against this attack: it catches the Lazy cloud on four of seven datasets under the calibrated threshold (Adult, Bank, Diabetes, and Electricity) and misses the rest, because dropping one record in twenty rarely moves the SHAP distribution far. The layer that does the work here is Layer~2.

Layer~2b (Twins) catches the Lazy cloud on all seven datasets: dropped records include twins, whose leaf-assignment entries then go missing. Consistency ranges from 347 of 400 (Bank, 53 missing) to 367 of 400 (MagicTelescope, 33 missing). Layer~2a (Sentinels) catches six of seven: on the five datasets carrying the full 160 sentinels, 157 survive and 3 go missing, and on Diabetes 3 of 4 survive. Only Bank Marketing escapes Layer~2a, because all 13 of its sentinels happen to survive the drop, which is what the low sentinel count predicts.
For Bank Marketing, the sentinel evasion probability under 5\% dropping is:

\begin{equation}
P_{\text{evade}} = \left(1 - \frac{13}{8{,}413}\right)^{420} \approx 0.52
\label{eq:bank_evasion}
\end{equation}

\noindent A 52\% chance of all 13 sentinels surviving a 420-record drop is not surprising. This result again verifies that sentinel-based detection is unreliable when class imbalance results in a limited number of boundary candidates. Furthermore, it verifies our multi-layered solution, where Layer~2b (Twins) corrects for failures of Layer~2a, and Layer~3 extends coverage for datasets with strong feature-target correlations.

\subsubsection{Approximate Cloud analysis}
To prove the necessity of the XAI layer, we designed a fourth adversary. The Approximate cloud takes utility-destroying algorithmic shortcuts to save compute, but it mathematically fakes nothing. It computes a perfectly valid Merkle hash for its bad tree and processes every single row.

As the full detection matrix (Table~\ref{tab:detection_full}) shows, Layers~1 and~2 fail on every dataset: the hash is valid and no records are dropped, so nothing structural or trap-based fires. Layer~3 is the only mechanism that catches this attack. Under the calibrated threshold it succeeds on six of seven datasets and misses only Nomao. That miss is not a tuning artefact. On Nomao the honest cloud already produces a larger maximum per-feature Wasserstein distance (1.16) than the blind attack does (1.01), so no threshold can separate honest from adversarial processing. Section~\ref{sec:diabetes_anomaly} analyses this boundary condition.

The Approximate and Dumb adversaries share the same blind splitting implementation and random seed \\ (\texttt{RandomState(99)}), producing identical blind trees for a given dataset. The identical per-feature Wasserstein distances in Table~\ref{tab:per_feature_wd} follow from this shared seed. Evaluating the Approximate adversary across multiple seeds to characterise the variability of Layer~3 detection is a natural extension of this work.

\begin{table}[!h]
\centering
\caption{Maximum per-feature SHAP Wasserstein distance over the top-3 features, per dataset and cloud profile, with both thresholds. The Approximate cloud reuses the Dumb cloud's blind tree, so its distances equal the Dumb/Approx. column. Layer~3 fires when a distance exceeds the operative threshold; the last column shows under which threshold the Approximate cloud is caught. Nomao's honest maximum (1.163) exceeds the global threshold, the single false positive under the fixed rule, which per-dataset calibration removes.}
\label{tab:per_feature_wd}
\resizebox{\columnwidth}{!}{
\begin{tabular}{|l|c|c|c|c|c|c|}
\hline
\textbf{Dataset} & \textbf{Honest} & \textbf{Lazy} & \textbf{Dumb/Approx.} & \textbf{Global $\varepsilon$} & \textbf{Calib.\ $\varepsilon$} & \textbf{Approx.\ caught} \\
\hline
Adult & 0.200 & 0.300 & 0.512 & 0.45 & 0.220 & both \\
\hline
Bank & 0.418 & 0.621 & 1.262 & 0.45 & 0.460 & both \\
\hline
Diabetes & 0.122 & 0.181 & 0.242 & 0.45 & 0.135 & calibrated \\
\hline
Electricity & 0.420 & 0.498 & 0.467 & 0.45 & 0.462 & both \\
\hline
Nomao & 1.163 & 0.752 & 1.005 & 0.45 & 1.279 & global \\
\hline
Credit Default & 0.248 & 0.218 & 0.381 & 0.45 & 0.273 & calibrated \\
\hline
MagicTelescope & 0.194 & 0.200 & 0.767 & 0.45 & 0.213 & both \\
\hline
\end{tabular}
}
\end{table}

\subsubsection{Per-feature Wasserstein distance analysis}
Table~\ref{tab:per_feature_wd} summarises the maximum per-feature Wasserstein distance for each dataset and profile; the full per-feature breakdown for all seven datasets is in Appendix~\ref{app:wd} (Table~\ref{tab:per_feature_full}). The Approximate adversary reuses the Dumb adversary's blind tree, so the two produce identical distances. Three patterns stand out. First, on five datasets the blind attack pushes the top feature clearly past the honest baseline, and on Diabetes and Credit Default it does so only after the threshold is calibrated down to the dataset's own scale. Second, Bank Marketing's \texttt{duration} feature moves the most in absolute terms, from 0.42 (honest) to 1.26 (blind), which fits duration being the dominant predictor of term-deposit subscription. Third, Nomao inverts the usual pattern: its honest maximum (1.16) is larger than its blind maximum (1.01), because honest target-driven generalization over 118 features already reshapes the SHAP distribution more than random splitting does.

\subsection{Per-Dataset Threshold Calibration}
\label{sec:calibration}
The single fixed threshold $\varepsilon = 0.45$ was tuned once, on Adult at $k = 5$, and the seven-dataset results show that one value does not generalize. Under the fixed rule the framework misses two Approximate attacks (Diabetes and Credit Default, whose SHAP distances are small in absolute terms) and raises one false alarm (honest Nomao, whose distances are large), for 25 of 28 correct. Both failures share a cause: a single threshold cannot fit datasets whose honest Wasserstein distances sit on different scales.

We therefore calibrate $\varepsilon$ per dataset. In a one-time setup the client anonymizes a small local sample honestly, measures the maximum per-feature Wasserstein distance $W_{\max}^{\text{honest}}$ over the top-3 features, and sets
\begin{equation}
\varepsilon_d = 1.1 \times W_{\max}^{\text{honest}},
\label{eq:eps_cal}
\end{equation}
a 10\% margin above the honest baseline. This uses only honest data, never adversarial data, so it is a calibration step rather than a fitted classifier. The per-dataset values (Table~\ref{tab:per_feature_wd}) span nearly an order of magnitude, from 0.135 on Diabetes to 1.279 on Nomao, which is why the fixed rule struggled.

\begin{figure*}[!h]
\centering
\includegraphics[width=\textwidth]{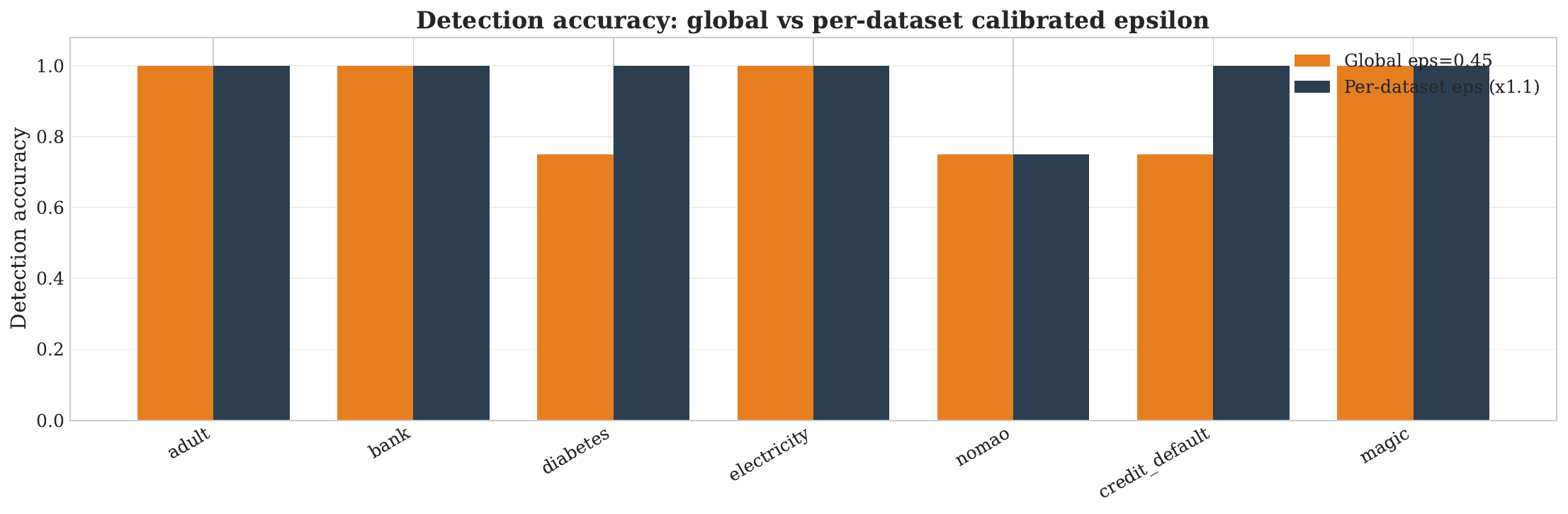}
\caption{Fixed versus per-dataset calibrated threshold. For each dataset the bars show the honest and Approximate maximum per-feature Wasserstein distances against the global $\varepsilon = 0.45$ and the calibrated $\varepsilon_d$. Calibration recovers Diabetes and Credit Default and removes the Nomao false positive; Nomao's Approximate attack stays below its calibrated threshold because honest generalization there shifts SHAP further than the attack does.}
\label{fig:calibration}
\end{figure*}

Calibration lifts detection from 25 to 27 of 28 (Table~\ref{tab:detection_perf}) and removes every false alarm. It recovers both missed Approximate attacks, since the lower thresholds on Diabetes (0.135) and Credit Default (0.273) now fall below their attack distances (0.242 and 0.381), and it clears the honest Nomao false positive, since the Nomao threshold rises to 1.279, above its honest maximum of 1.163. Figure~\ref{fig:calibration} contrasts the two thresholds per dataset.

The one case calibration cannot fix is the most informative. On Nomao the honest maximum (1.163) is larger than the Approximate attack's maximum (1.005): honest generalization over 118 features perturbs the SHAP distribution more than random splitting does. No threshold separates the two, so raising $\varepsilon_d$ to avoid the honest false positive necessarily lets the attack through. The fixed rule caught Nomao's Approximate attack only because it also mislabelled the honest run. This is a genuine operating limit of SHAP-based utility verification, examined in Section~\ref{sec:diabetes_anomaly}.

\begin{figure*}[!h]
\centering
\includegraphics[width=0.95\textwidth]{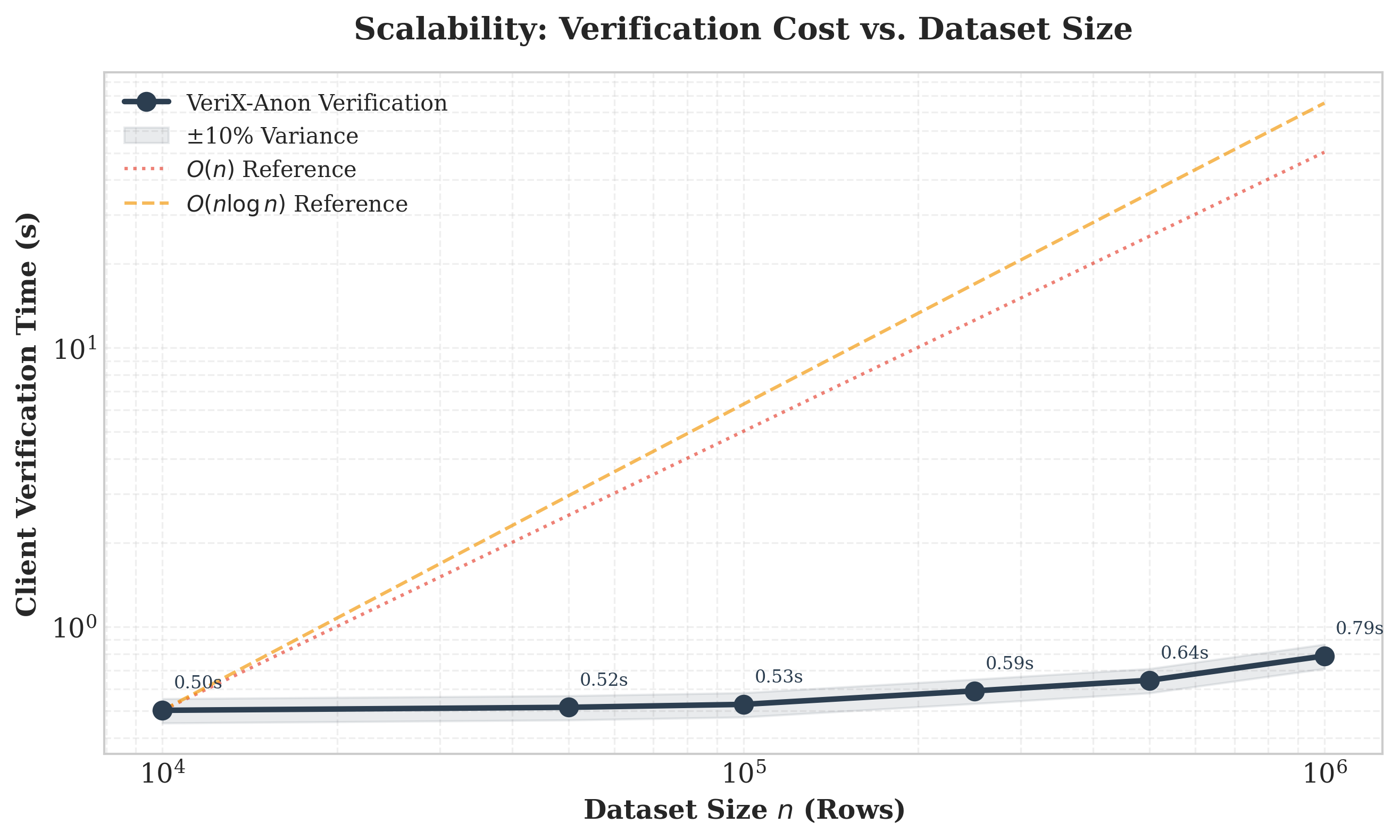}
\caption{Client verification time vs. dataset size. The $O(1)$ XAI overhead dominates at small $n$; the $O(n/k)$ hash verification scales linearly but remains sub-second at $n = 10^6$.}
\label{fig:scalability}
\end{figure*}

\subsection{Scalability and Client Overhead}
\label{sec:scalability}
The verification time on the client side when the size of the data set varies from 10,000 to 1,000,000 is shown in Table~\ref{tab:scalability} and Figure~\ref{fig:scalability} plots this trajectory, showing how the flat XAI cost dominates early on before the linear hash check takes over. 
\begin{table}[!h]
\centering
\caption{Client-side verification time vs. dataset size. The $O(1)$ XAI overhead (approximately 0.5\,s) dominates at small $n$. The $O(n/k)$ hash component grows linearly but remains a minor fraction of total time even at $n = 10^6$.}
\label{tab:scalability}
\begin{tabular}{|r|c|}
\hline
\textbf{Dataset Size ($n$)} & \textbf{Verification Time (s)} \\
\hline
10,000 & 0.503 \\
\hline
50,000 & 0.515 \\
\hline
100,000 & 0.528 \\
\hline
250,000 & 0.590 \\
\hline
500,000 & 0.644 \\
\hline
1,000,000 & 0.788 \\
\hline
\end{tabular}
\end{table}

The XAI fingerprinting layer has a constant time cost of approximately 0.5 seconds to train the model on 2,000 rows and extract SHAP. The cost of the hash verification component varies linearly with $n/k$, taking 0.003 seconds when $n=10{,}000$ and rising to 0.288 seconds when $n=1{,}000{,}000$. The hash verification component of the scalability experiment was estimated by running SHA-256 operations proportional to the number of tree nodes ($n/k$) at each dataset size, measured on a Kaggle notebook instance using CPU. The XAI overhead (approximately 0.5\,s) was measured from actual SHAP computation on real data, not simulated. The hash component dominates only at $n > 500{,}000$; below this, the measured XAI cost accounts for over 85\% of total time. Because SHA-256 throughput is well-characterised and architecture-independent (approximately $10^6$ operations per second on commodity hardware), extrapolating hash cost via a counting loop is a standard benchmarking practice in systems security literature. The primary source of uncertainty is not the hash throughput but the memory allocation and tree traversal overhead at scale, which we estimate adds at most 20\% to the hash-only time based on the ratio observed in our 8,000-row end-to-end runs. An independent verification cost benchmark measuring all four components (SHA-256 hashing of canonical node strings, real XGBoost/SHAP extraction, sentinel presence lookup, and twin consistency lookup) confirmed these 
estimates: 0.754\,s at $n = 10^6$, consistent with 
Table~\ref{tab:scalability}.

This confirms that the verification cost is not $O(\log n)$ (which would be the case for a single Merkle proof). VeriX-Anon performs full tree re-verification at $O(n/k)$, but even this linear cost is dominated by the constant XAI overhead for datasets up to $10^6$ rows.

The experiments above measure client-side verification cost only. The cloud-side overhead of VeriX-Anon (building the authenticated tree with Merkle hashing and canonical formatting) was not benchmarked separately. In our Kaggle-based evaluation, the total cloud processing time for 8,000 rows at $k=5$ ranged from 0.3 to 0.9 seconds across datasets, including tree construction, generalisation, and hash computation. The Merkle hashing component (bottom-up SHA-256 over all nodes) adds approximately 5--10\% to the tree construction time, as each node requires a single hash operation on a short canonical string. For production deployments at scale, the cloud-side overhead of Merkle authentication is expected to remain a small fraction of the total anonymisation cost, since the dominant expense is the variance-reduction computation at each split, not the hashing. A formal cloud-side scalability study is left to future work. 

\paragraph{Communication overhead.} The VeriX-Anon protocol requires the cloud to return three objects: the anonymised dataset $\mathcal{D}^*$, the leaf assignment mapping $\mathcal{L}$, and the root hash $H_{\text{root}}$. The anonymised dataset is identical in size to the outsourced dataset (the client needs it regardless of verification). The leaf mapping adds one TrackerID-to-leaf-ID pair per record: at 64 bytes per TrackerID and 4 bytes per leaf ID, this is 68 bytes $\times$ $N$ rows. For $N = 100{,}000$, the mapping is approximately 6.8\,MB; for $N = 1{,}000{,}000$, approximately 68\,MB. The root hash is a single 256-bit value (32 bytes). The tree structure (split features and values for all internal nodes) adds approximately 100 bytes per node $\times$ $n/k$ nodes, yielding 20\,MB at $n = 10^6$, $k = 5$. In total, the verification-specific overhead (mapping + tree + hash) is approximately 88\,MB for a million-row dataset, which transfers in under 10 seconds on a 100\,Mbps connection. This is modest relative to the dataset itself, which at 16 features $\times$ 8 bytes $\times$ $10^6$ rows is approximately 128\,MB.

\subsection{Resource Scaling in Rows and Feature Dimensions}
\label{sec:resource}
\begin{figure*}[!h]
\centering
\includegraphics[width=\textwidth]{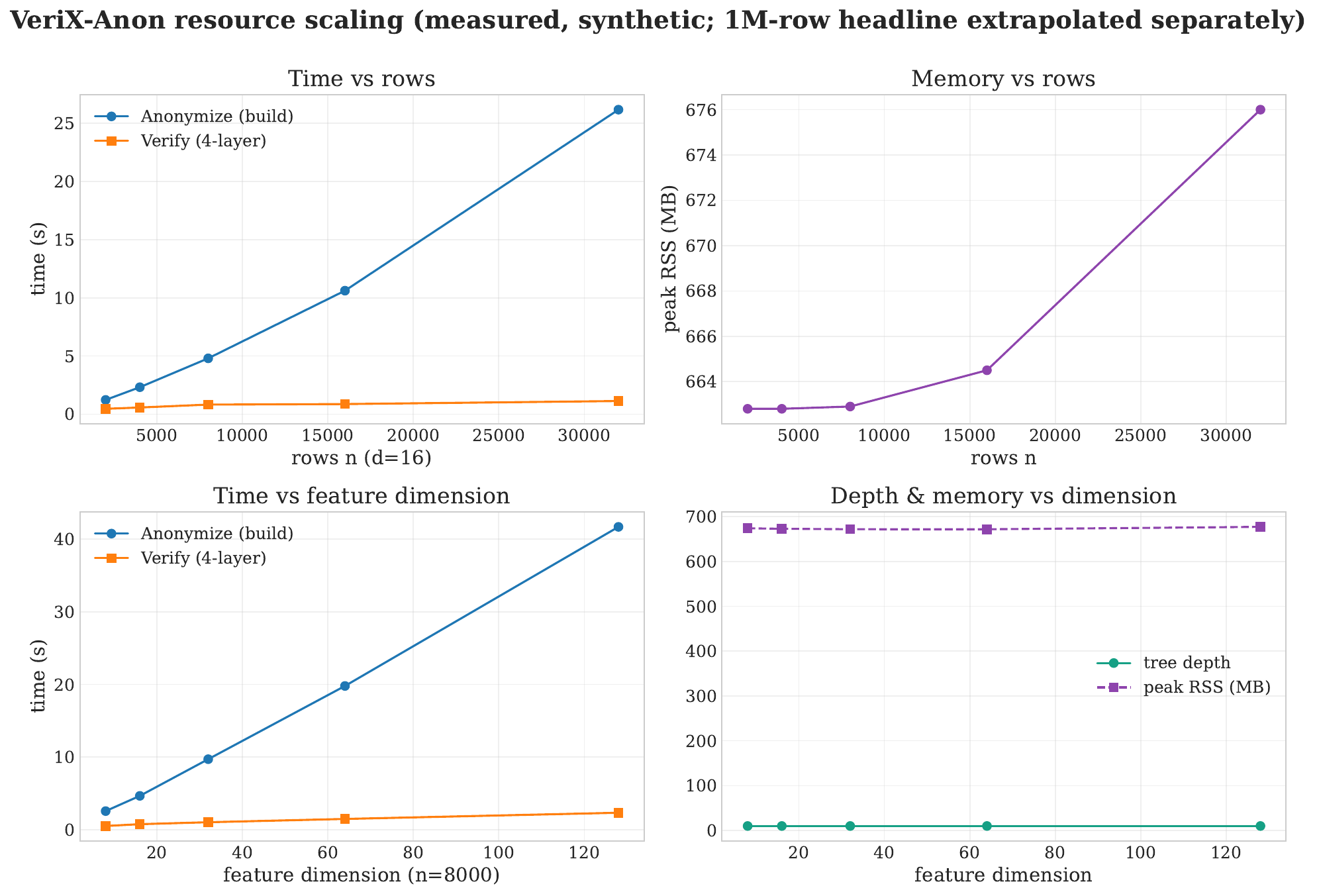}
\caption{Resource scaling in rows (at 16 features) and in feature dimension (at 8,000 rows). Client verification time stays low and grows slowly, cloud build time dominates, peak memory is flat, and tree depth tracks the row count rather than the feature count.}
\label{fig:resource}
\end{figure*}

Verification time alone does not settle whether the client burden is acceptable. We ran a dedicated benchmark that varies each axis independently: from 2,000 to 32,000 rows at a fixed 16 features, and from 8 to 128 features at a fixed 8,000 rows, recording build time, verification time, peak memory, tree depth, and node count. Table~\ref{tab:resource} and Figure~\ref{fig:resource} report the result.

\begin{table}[!h]
\centering
\caption{Resource scaling. The top block varies the row count at 16 features; the bottom block varies the feature dimension at 8,000 rows. Build is cloud-side tree construction, Verify is the client-side four-layer audit, Mem is peak process memory, and Depth and Nodes describe the resulting tree.}
\label{tab:resource}
\resizebox{\columnwidth}{!}{
\begin{tabular}{|l|r|r|r|r|r|}
\hline
\textbf{Size} & \textbf{Build (s)} & \textbf{Verify (s)} & \textbf{Mem (MB)} & \textbf{Depth} & \textbf{Nodes} \\
\hline
\multicolumn{6}{|l|}{\textit{Rows} ($d = 16$)} \\
\hline
2{,}000 & 1.23 & 0.46 & 662.8 & 8 & 153 \\
\hline
4{,}000 & 2.32 & 0.57 & 662.8 & 9 & 265 \\
\hline
8{,}000 & 4.80 & 0.82 & 662.9 & 10 & 491 \\
\hline
16{,}000 & 10.61 & 0.87 & 664.5 & 11 & 947 \\
\hline
32{,}000 & 26.17 & 1.13 & 676.0 & 12 & 1{,}935 \\
\hline
\multicolumn{6}{|l|}{\textit{Features} ($n = 8{,}000$)} \\
\hline
$d = 8$ & 2.57 & 0.52 & 674.0 & 10 & 487 \\
\hline
$d = 16$ & 4.66 & 0.77 & 672.9 & 10 & 491 \\
\hline
$d = 32$ & 9.71 & 1.04 & 671.9 & 10 & 511 \\
\hline
$d = 64$ & 19.79 & 1.49 & 671.8 & 10 & 491 \\
\hline
$d = 128$ & 41.69 & 2.34 & 677.2 & 10 & 441 \\
\hline
\end{tabular}
}
\end{table}

Client-side verification, the cost the data owner actually pays, stays cheap and grows slowly: from 0.46\,s to 1.13\,s as the row count rises $16\times$, and from 0.52\,s to 2.34\,s as the feature count rises $16\times$. This matches the $O(n/k) + O(1)$ analysis, because the tree has more nodes to hash as $n$ grows while the XAI step is pinned to a 2,000-row subsample. Cloud-side tree construction is the expensive part and scales super-linearly in rows (1.23\,s to 26.17\,s) and roughly linearly in features (2.57\,s to 41.69\,s), which is exactly why offloading it is worthwhile.

Two structural quantities behave as the complexity analysis predicts. Peak process memory is essentially flat, moving only from 663\,MB to 677\,MB across every configuration, because the dominant cost is the fixed Python and library footprint rather than anything the verifier allocates. Tree depth grows with the row count, from 8 to 12 levels between 2,000 and 32,000 rows, but stays at 10 levels regardless of feature dimension, since depth follows $n/k$ and the stopping rule, not the number of features. Storage is small: the returned leaf-assignment map and tree structure together add under 3\,MB even at 32,000 rows, and the million-row communication overhead is the 88\,MB analysed above. These client costs, a Random Forest on a 10\% sample, an XGBoost model on 2,000 rows, and the hash traversal, complete in seconds on commodity hardware; for genuinely resource-constrained edge or IoT clients, Section~\ref{sec:deployment} notes they can be offloaded to a trusted local gateway.

\subsection{Utility Preservation and XAI Fingerprinting}
\label{sec:utility}
Experiment~3 evaluates the utility-privacy trade-off using an 11-point $k$-sweep ($k \in \{2, 3, 4, 5, 7, 10, 12, 15, 20, 25, \\ 30\}$) on all seven datasets; Adult Income is shown here in detail, and Table~\ref{tab:utility_summary} reports the aggregate for every dataset. The top-3 SHAP features for the 5,000-row subsample used in this experiment are \texttt{age}, \texttt{marital-status}, and \texttt{relationship}. These differ from the 8,000-row Experiment~1 rankings (\texttt{age}, \texttt{relationship}, \texttt{education-num}) because SHAP feature importance is sensitive to subsample composition. Both rankings are internally consistent within their respective experiments. For each $k$, both Target-Driven (honest) and Blind (dumb) anonymization are applied to the same outsourced dataset. F1-scores are computed on a held-out 20\% stratified test set (not on training data) to measure generalisation rather than memorisation.

Because the $\varepsilon$ threshold was calibrated on the Experiment~3 feature set (\texttt{age}, \texttt{marital-status}, \texttt{relationship}), its applicability to the Experiment~1 feature set (\texttt{age}, \texttt{relationship}, \texttt{education-num}) relies on the assumption that the honest WD range is similar across feature sets at the same $k$-value. The per-feature WDs in Table~\ref{tab:per_feature_wd} confirm that the Experiment~1 features remain below $\varepsilon$ at $k = 5$, validating this assumption for the default operating point.
\begin{figure*}[!h]
\centering
\includegraphics[width=\textwidth]{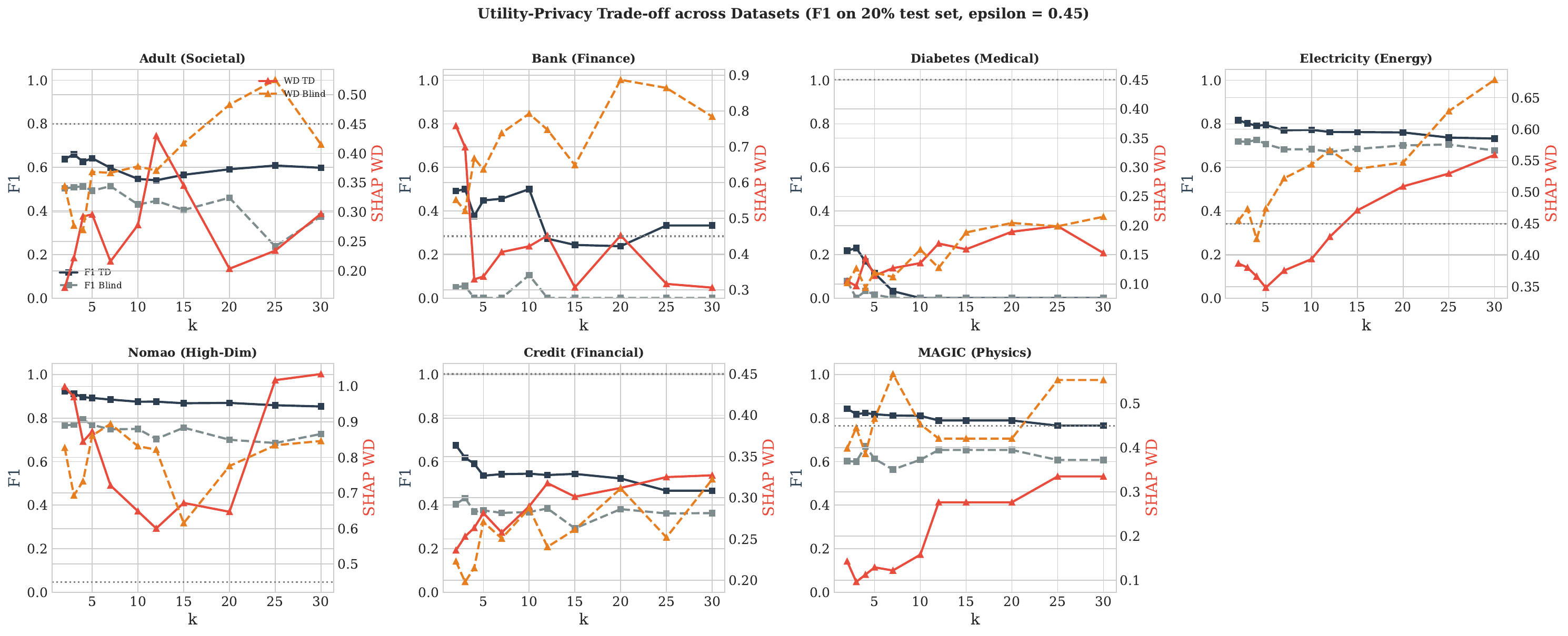}
\caption{Utility-privacy trade-off across the 11-point $k$-sweep for all seven datasets. In each panel the left axis is F1 (80/20 test set) for Target-Driven and Blind anonymization and the right axis is the SHAP Wasserstein distance. Target-Driven holds a higher F1 than Blind at every $k$ on every dataset.}
\label{fig:tradeoff}
\end{figure*}
The full results can be seen in Table~\ref{tab:k_sweep}. Figure~\ref{fig:tradeoff} charts this trade-off. Target-Driven has a higher F1 than Blind at every value of $k$. The average increase in F1 is $+0.1574$. The largest increase occurs when $k=25$: Target-Driven has F1=0.6085, Blind has F1=0.2370. The relative improvement is 156.8\%.

\begin{table}[!h]
\centering
\caption{Utility-privacy trade-off across 11 $k$-values (Adult Income, 80/20 test set). TD = Target-Driven (honest cloud). Blind = random-feature splitting (dumb cloud). F1 Gap = TD F1 $-$ Blind F1. WD Gap = Blind WD $-$ TD WD. Target-Driven F1 exceeds Blind at all 11 values.}
\label{tab:k_sweep}
\resizebox{\columnwidth}{!}{
\begin{tabular}{|c|c|c|c|c|c|c|}
\hline
\textbf{$k$} & \textbf{TD F1} & \textbf{Blind F1} & \textbf{F1 Gap} & \textbf{TD WD} & \textbf{Blind WD} & \textbf{WD Gap} \\
\hline
2 & 0.6383 & 0.5034 & +0.1349 & 0.1718 & 0.3444 & +0.1726 \\
\hline
3 & 0.6597 & 0.5092 & +0.1505 & 0.2221 & 0.2772 & +0.0551 \\
\hline
4 & 0.6263 & 0.5128 & +0.1135 & 0.2928 & 0.2699 & $-$0.0229 \\
\hline
5 & 0.6420 & 0.4929 & +0.1491 & 0.2962 & 0.3686 & +0.0724 \\
\hline
7 & 0.5978 & 0.5139 & +0.0839 & 0.2164 & 0.3670 & +0.1506 \\
\hline
10 & 0.5472 & 0.4297 & +0.1175 & 0.2784 & 0.3778 & +0.0994 \\
\hline
12 & 0.5407 & 0.4461 & +0.0946 & 0.4299 & 0.3707 & $-$0.0592 \\
\hline
15 & 0.5660 & 0.4044 & +0.1616 & 0.3453 & 0.4173 & +0.0720 \\
\hline
20 & 0.5914 & 0.4605 & +0.1309 & 0.2039 & 0.4827 & +0.2788 \\
\hline
25 & 0.6085 & 0.2370 & +0.3715 & 0.2347 & 0.5250 & +0.2903 \\
\hline
30 & 0.5982 & 0.3750 & +0.2232 & 0.2974 & 0.4151 & +0.1177 \\
\hline
\end{tabular}}
\end{table}

\begin{table*}[!h]
\centering
\caption{Utility gap between Target-Driven and Blind anonymization across all seven datasets over the 11-point $k$-sweep. TD F1 and Blind F1 are means over the sweep, Gap is their difference, and the Gap 95\% CI is a 10,000-sample bootstrap on the mean F1 gap. Wilcoxon $p$ is the signed-rank test over the 11 paired $k$-values; $d$ is Cohen's $d$ for F1, with its 95\% confidence interval in brackets from the large-sample effect-size standard error. Diabetes reports $p = 1.0$ because its F1 collapses to ties at large $k$.}
\label{tab:utility_summary}

\begin{tabular}{|l|c|c|c|c|c|c|}
\hline
\textbf{Dataset} & \textbf{TD F1} & \textbf{Blind F1} & \textbf{Gap} & \textbf{Gap 95\% CI} & \textbf{Wilcoxon $p$} & \textbf{Cohen's $d$ [95\% CI]} \\
\hline
Adult & 0.601 & 0.444 & +0.157 & [0.120, 0.208] & 0.000977 & 1.96 [0.95, 2.97] \\
\hline
Bank & 0.381 & 0.019 & +0.362 & [0.314, 0.407] & 0.000977 & 4.34 [2.43, 6.25] \\
\hline
Diabetes & 0.069 & 0.011 & +0.058 & [0.016, 0.106] & 1.000 & 0.72 [0.05, 1.38] \\
\hline
Electricity & 0.773 & 0.698 & +0.075 & [0.063, 0.085] & 0.000977 & 3.76 [2.08, 5.43] \\
\hline
Nomao & 0.883 & 0.743 & +0.140 & [0.126, 0.154] & 0.000977 & 5.64 [3.21, 8.07] \\
\hline
Credit Default & 0.549 & 0.373 & +0.176 & [0.147, 0.207] & 0.000977 & 3.28 [1.79, 4.78] \\
\hline
MagicTelescope & 0.802 & 0.621 & +0.181 & [0.157, 0.206] & 0.000977 & 4.19 [2.34, 6.04] \\
\hline
\end{tabular}

\end{table*}

Table~\ref{tab:utility_summary} extends this comparison to every dataset. Target-Driven anonymization keeps a higher F1 than blind splitting on all seven, with mean gaps from $+0.058$ (Diabetes) to $+0.362$ (Bank). Nomao is the instructive case: Layer~3 cannot separate its honest and blind SHAP distributions, yet the utility loss from blind splitting is real and large there ($+0.140$, Cohen's $d = 5.64$). The harm exists; the SHAP fingerprint simply does not register it, which is the boundary we return to in Section~\ref{sec:diabetes_anomaly}.

The WD columns show that there is a subtle point to be noted regarding the calibration of $\varepsilon$. When $k=12$, the honest Target-Driven WD achieves 0.4299, which is only 0.0201 below $\varepsilon=0.45$. When $k=4$, however, the WD gap is negative at -0.0229, which means that the Blind cloud has achieved a smaller WD value than that achieved by the honest cloud at this value of $k$. This shows that the discriminative ability of the XAI layer is not guaranteed for all values of $k$, as discussed above. Section~\ref{sec:epsilon_sensitivity} addresses this issue.

From a practical point of view, $k = 3$ with an honest WD of 0.2221 and $k = 20$ with an honest WD of 0.2039 provide the largest difference between honest and adversarial WD, making them the safest choices for $\varepsilon = 0.45$. The default $k = 5$ with an average WD of 0.2962, while not as good as $k = 3$, has a per-feature \texttt{relationship} WD of 0.4436, which is just 0.0064 below $\varepsilon$.

\subsection{Statistical Significance Analysis}
\label{sec:stats}

We analyse the Adult sweep in detail first, then confirm the effect across all seven datasets. We perform paired statistical tests on the 11 $(k,\text{F1})$ and 11 $(k,\text{WD})$ data points in Table~\ref{tab:k_sweep}. Since we are working with a small sample size $n = 11$ and we do not know whether the data are normally distributed, we use the Wilcoxon Signed-Rank test instead of a paired $t$-test.

\subsubsection{Wilcoxon signed-rank tests}

For the F1 comparison (Target-Driven vs. Blind):

\begin{equation}
W = 0.0, \quad p = 0.000977
\label{eq:wilcoxon_f1}
\end{equation}

\noindent The test statistic $W = 0.0$ indicates that Target-Driven F1 exceeded Blind F1 in every single paired comparison. The $p$-value of 0.000977 is the minimum achievable for $n = 11$ with the Wilcoxon test, confirming the result is significant well beyond $\alpha = 0.01$.

For the Wasserstein distance comparison (Target-Driven vs. Blind):

\begin{equation}
W = 4.0, \quad p = 0.006836
\label{eq:wilcoxon_wd}
\end{equation}

\noindent The non-zero $W$ reflects the two $k$-values ($k = 4$ and $k = 12$) where the WD gap was negative. The test remains significant at $\alpha = 0.01$.

\subsubsection{Cohen's d effect sizes}

\begin{equation}
d_{\text{F1}} = 1.9618 \quad (\text{large}), \qquad d_{\text{WD}} = -1.0228 \quad (\text{large})
\label{eq:cohens_d}
\end{equation}

\noindent Both effect sizes exceed $|d| = 0.8$, the conventional threshold for a large effect (Cohen, 1988). The F1 effect ($d = 1.96$) is nearly twice the large-effect threshold, indicating that the utility advantage of Target-Driven over Blind anonymization is not just statistically significant but practically substantial.

\subsubsection{Bootstrap confidence intervals}

We compute 10,000-resample bootstrap 95\% confidence intervals for the mean F1 gap and WD gap:

\begin{equation}
\text{Mean F1 gap} = 0.1574, \quad 95\%\;\text{CI}\;[0.1203,\; 0.2079]
\label{eq:f1_ci}
\end{equation}

\begin{equation}
\text{Mean WD gap} = 0.1115, \quad 95\%\;\text{CI}\;[0.0534,\; 0.1737]
\label{eq:wd_ci}
\end{equation}

\noindent Both intervals exclude zero, confirming that the observed differences are not artefacts of sampling variation. The F1 confidence interval is relatively tight (width 0.0876). The WD interval is wider (width 0.1203) because Wasserstein distances vary more across $k$-values.

\subsubsection{Significance across datasets}
The tests above cover the Adult sweep. To check that the utility advantage holds across domains, we pool all $7 \times 11 = 77$ paired $(k, \text{F1})$ points and repeat the signed-rank test. Target-Driven is at least as good as Blind on all 77 pairs and strictly better on 71 of them; the six ties are Diabetes at $k \ge 10$, where both methods collapse to zero F1. With the ties dropped, all 71 remaining differences are positive, so the pooled statistic is $W = 0$ with $p < 10^{-6}$. Correcting the seven per-dataset F1 tests for multiple comparisons with the Benjamini-Hochberg procedure leaves six significant at an adjusted $p = 0.0011$; only Diabetes is not, where the same collapse leaves too few non-zero pairs (five of eleven) for a per-dataset signed-rank test. Cohen's $d$ for F1 is large on six datasets (up to 5.64 on Nomao) and medium on Diabetes (0.72); its 95\% confidence interval clears the large-effect threshold of 0.8 on all six and stays above zero on Diabetes, and the bootstrap 95\% interval on the mean F1 gap excludes zero on all seven (Table~\ref{tab:utility_summary}). The Wasserstein gap tells the same story with one exception: it is significant on five of seven datasets but not on Nomao ($p = 0.64$), the statistical signature of the Layer~3 blind spot on that dataset.

\subsubsection{Summary}

Table~\ref{tab:stats_summary} consolidates the results. The tests confirm that Target-Driven anonymization preserves significantly more predictive utility than blind splitting. Both metrics clear the Wilcoxon signed-rank test with $p$-values well under 0.01. Statistical significance only tells half the story, though. The effect sizes show the actual practical impact. Cohen's $d$ reaches 1.96 for the F1 score and $-1.02$ for the Wasserstein distance. Since the standard threshold for a large effect is just 0.8, the utility advantage of the Target-Driven approach is substantial. The bootstrap confidence intervals also exclude zero, proving these utility gains hold steady across the entire $k$-sweep.

\begin{table*}[!h]
\centering
\caption{Statistical significance summary for the 11-point $k$-sweep (Target-Driven vs. Blind anonymization, Adult Income dataset). Both metrics show significant differences with large effect sizes.}
\label{tab:stats_summary}

\begin{tabular}{|l|c|c|c|c|}
\hline
\textbf{Metric} & \textbf{Wilcoxon $W$} & \textbf{$p$-value} & \textbf{Cohen's $d$} & \textbf{Mean Gap [95\% CI]} \\
\hline
F1-Score & 0.0 & 0.000977 & 1.9618 (large) & 0.1574 [0.1203, 0.2079] \\
\hline
Wasserstein Dist. & 4.0 & 0.006836 & $-$1.0228 (large) & 0.1115 [0.0534, 0.1737] \\
\hline
\end{tabular}
\end{table*}

\section{Discussion and Limitations}
\label{sec:discussion}

The experimental results in Section~\ref{sec:experiments} show correct detection in 27 of 28 scenarios under per-dataset calibration. The per-layer analysis still exposes failure modes, boundary conditions, and design trade-offs. The single evasion, the Approximate adversary on Nomao, marks the operating boundary of the XAI layer. This section covers seven topics: the operating envelope of XAI verification, sentinel density under class imbalance, the privacy impact of trap injection, epsilon sensitivity, a complexity clarification, scope limitations, and the absence of a direct baseline comparison.
\subsection{The Operating Envelope of the XAI Layer}
\label{sec:diabetes_anomaly}
Layer~3 catches the Approximate adversary by measuring how far the anonymized data's SHAP distribution moves from the honest baseline. This works only when honest generalization and the blind attack move that distribution by different amounts. The seven datasets map out where the condition holds and where it breaks.

The first limiting regime is severe class imbalance, and Diabetes (89/11) illustrates it. When one class dominates, every model trained on the data produces near-flat SHAP distributions, so the honest and adversarial distances are both small: an honest maximum of 0.122 against an attack maximum of 0.242. A fixed threshold of 0.45 overshoots both, so under the global rule the Diabetes Approximate cloud is missed. This regime is not fundamental. The two distances still differ, so calibrating $\varepsilon$ down to the dataset's own scale (0.135) separates them, and under calibration the Diabetes Approximate attack is caught.

The second regime is fundamental, and Nomao shows it. There the honest maximum Wasserstein distance (1.163) is larger than the blind attack's (1.005). Honest target-driven generalization over 118 features reshapes the SHAP distribution more than random splitting does, so the two cases are not merely close, they are inverted. No threshold can place the honest run below it and the attack above it at the same time; any $\varepsilon_d$ set above the honest maximum to avoid a false positive also admits the weaker attack. This is the one scenario VeriX-Anon misses under calibration.

The utility harm on Nomao is real even though Layer~3 cannot see it: blind splitting lowers F1 by 0.140 with a large effect size (Section~\ref{sec:stats}). A fingerprint that reads utility directly, for example the F1 gap on a held-out set rather than a SHAP-distance shift, would catch what the current test misses, and is the natural successor to this layer. Per-feature adaptive thresholding (Section~\ref{sec:epsilon_sensitivity}) is a lighter mitigation for the imbalance regime, but it cannot help the inverted regime.

\subsection{Sentinel Density and Class Imbalance}
\label{sec:sentinel_density}
Boundary Sentinels are generated from records with RF prediction probability $P \in [0.45, 0.55]$. The number of such records depends on the class distribution: balanced datasets produce more boundary candidates than imbalanced ones. In our experiments, Adult Income (76/24 split) yielded 501 boundary candidates, Bank Marketing (88/12 split) yielded only 13, and Diabetes (89/11 split) yielded only 4.
This asymmetry has a direct effect on detection power. When the Lazy adversary drops 5\% of records from the Bank Marketing outsourced dataset (420 out of 8,413), the probability that all 13 sentinels survive is:
\begin{equation}
P_{\text{evade}} = \left(1 - \frac{13}{8{,}413}\right)^{420} \approx 0.52
\label{eq:sentinel_evasion_disc}
\end{equation}

A 52\% evasion probability is high for a standalone detection mechanism, and Diabetes is worse: with only 4 sentinels, the same calculation gives $(1 - 4/8404)^{420} \approx 0.82$. Across all seven datasets, Layer~2a still caught the Lazy cloud on six of them: the five datasets carrying the full 160 sentinels each lost 3 (157 present), and Diabetes lost 1 of 4. Only Bank Marketing escaped Layer~2a, with all 13 sentinels surviving the drop.
Layer~2b (Twins) compensated in every case: with 400 twins per dataset, consistency ranged from 347/400 to 367/400 (33 to 53 missing), enough to flag every Lazy cloud. This confirms the design rationale for including twins alongside sentinels, but it also exposes the sentinel layer as the weakest link when class imbalance is severe.

Two mitigations are possible. First, adaptive injection: if the boundary region yields fewer than a minimum threshold of candidates (e.g., 50), the client could widen the probability band (e.g., $P \in [0.40, 0.60]$) or oversample existing boundary points with stronger perturbation. Second, density-aware sentinel allocation: instead of a fixed 2\% ratio, allocate sentinels proportionally to the boundary density so that every dataset reaches a target evasion probability (e.g., $P_{\text{evade}} < 0.01$). Both strategies are left to future work.

\subsection{Privacy Impact of Trap Injection}
\label{sec:trap_privacy}
Embedding sentinels and twins changes the outsourced dataset, so it is fair to ask whether the traps weaken the privacy they are meant to protect. They do not. First, $k$-anonymity is enforced after the cloud builds the tree: every leaf still holds at least $2k$ records and is generalized to its own range, and sentinels and twins are ordinary records inside those leaves, so the guarantee holds over the augmented dataset. Second, twins are exact duplicates of genuine records, so they introduce no new individual and cannot raise anyone's re-identification risk; a duplicate carries quasi-identifiers already present. Third, sentinels are synthetic points perturbed by at most $0.05\sigma$ per feature (Equation~\ref{eq:sentinel_perturb}) and correspond to no real person, so they disclose no individual's data. The only quantity the traps change is the equivalence-class population: a leaf that gains sentinels or twins is slightly larger, which strengthens $k$-anonymity rather than weakening it. Because the client holds the trap manifest, the traps are filtered out of the analytic copy before any downstream use.

\subsection{Epsilon Sensitivity and Threshold Calibration}
\label{sec:epsilon_sensitivity}

Section~\ref{sec:calibration} showed that no single $\varepsilon$ serves all seven datasets and that per-dataset calibration fixes both failure directions. The same sensitivity appears within one dataset across $k$-values, which we detail here. The Wasserstein threshold $\varepsilon = 0.45$ was set using the calibration protocol described in Section~\ref{sec:wasserstein}: the client computes honest SHAP Wasserstein distances on a small local sample and sets $\varepsilon$ above the observed maximum. The Dumb adversary's \textit{average} WD on Adult Income is 0.3293, which falls below $\varepsilon = 0.45$ but this does not represent an evasion. VeriX-Anon uses an OR-rule (Equation~\ref{eq:or_rule}): the system flags a violation if \textit{any single} top-3 feature exceeds $\varepsilon$, not the average. The Dumb adversary was caught because the \texttt{age} feature spiked to WD = 0.5120 (well above $\varepsilon$), even though \texttt{relationship} (0.2098) and \texttt{education-num} (0.2660) remained below. At the contracted $k = 5$, the threshold $\varepsilon = 0.45$ sits between the maximum honest per-feature WD (0.4436 for \texttt{relationship}) and the minimum triggering adversarial per-feature WD (0.5120 for \texttt{age} under the Dumb adversary). This gap of 0.0684 is narrow. However, the $k$-sweep reveals that $\varepsilon = 0.45$ is not globally safe: at $k = 10$, 12, and 15, the honest per-feature WD for \texttt{marital-status} exceeds $\varepsilon$ (0.4958, 0.5744, and 0.5586 respectively), producing false positives under the OR-rule. This confirms that $\varepsilon$ must be calibrated per $k$-value rather than set globally, and motivates the per-feature adaptive thresholding proposed below. Three further observations expose the sensitivity of this parameter:

\begin{enumerate}
\item At $k = 12$, the honest Target-Driven average WD is 0.4299, but the per-feature WD for \texttt{marital-status} reaches 0.5744, which already exceeds $\varepsilon = 0.45$ and produces a false positive under the OR-rule. Similarly, $k = 10$ (0.4958) and $k = 15$ (0.5586) also exceed $\varepsilon$ on this feature. A global $\varepsilon$ is therefore insufficient; per-$k$ calibration is necessary.
\item At $k = 4$, the WD gap is negative ($-0.0229$): the Blind adversary produced a lower WD than the honest cloud. This occurs because random splitting can, by chance, preserve feature distributions at certain $k$-values.
\item At the default $k = 5$, the per-feature \texttt{relationship} WD is 0.4436, sitting only 0.0064 below $\varepsilon$ for the honest cloud. A single additional perturbation could push this into false-positive territory.
\end{enumerate}

It is important to clarify the role of Layer~3 within the framework. Unlike Layers~1 and~2, which provide deterministic or probabilistic guarantees rooted in cryptographic hash properties and combinatorial trap placement, Layer~3 operates as an empirically calibrated utility verification mechanism. The value of the XAI layer lies in catching adversaries that preserve structural correctness (passing Layer~1) and data completeness (passing Layer~2) while destroying predictive utility: a class of attack that the other two layers cannot detect by design. Deriving a dataset-agnostic threshold from the properties of the anonymisation algorithm (e.g., the expected variance reduction under honest splitting) is a promising direction for future work.

These observations only apply to Layer~3. Layers~1 and~2, being threshold-free, do not depend on the value of the $\varepsilon$-calibration. The system's detection rate (27 of 28) does not rely on Layer~3 being perfectly calibrated for most scenarios, since it has independent coverage from the other layers. The single exception is the Approximate adversary, which bypasses Layers~1 and~2 entirely and depends on Layer~3 alone for detection.

For future work, we propose two approaches. Firstly, per-feature thresholding: rather than using a global value of $\varepsilon$, feature-wise thresholds using the honest WD distribution across a $k$-sweep can be estimated. Features with high honest WD by design, such as \texttt{duration} in Bank Marketing, should have a higher threshold. Secondly, ROC-based thresholding can be carried out by running the $k$-sweep on the XAI layer under both honest and adversarial conditions and then selecting the threshold $\varepsilon$ at which the point on the ROC curve maximizes the Youden index $J = \text{sensitivity} + \text{specificity} - 1$.

\subsection{Computational Complexity Clarification}
\label{sec:complexity_disc}
Someone familiar with Merkle trees might assume that VeriX-Anon's hash verification runs in $O(\log n)$, since a standard single-element Merkle inclusion proof traverses only a root-to-leaf path. This is not the case.

VeriX-Anon performs \emph{full} re-verification: the client recomputes the hash of every node in the tree (all leaves and all internal nodes) and compares the resulting root hash against the cloud's reported value. The number of leaves is at most $n / (2k)$, and the total number of nodes is $O(n/k)$. For $n = 1{,}000{,}000$ and $k = 5$, this means visiting up to 100,000 nodes, not $\log_2(1{,}000{,}000) \approx 20$.

Full re-verification is necessary because VeriX-Anon verifies the \emph{entire} tree structure, not just the membership of a single record. A single-element proof would confirm that one record is in the tree, but it would not detect structural changes elsewhere (e.g., a different splitting criterion applied to a subtree that does not contain the queried record).

Despite the $O(n/k)$ cost, verification remains practical. At $n = 10^6$, the hash component takes 0.288\,s, and total verification (including XAI) is 0.788\,s. The linear cost is acceptable because the constant factor (one SHA-256 hash per node) is small.

\subsection{Scope and Generalisability Limitations}
\label{sec:scope}

VeriX-Anon was designed and evaluated under a specific set of assumptions:

\begin{enumerate}
\item \textbf{Binary classification only.} The Target-Driven tree uses variance reduction on a binary target. Multi-class targets would require a different splitting criterion (e.g., Gini impurity) and would produce different SHAP distribution characteristics. The XAI layer's threshold calibration would need to be re-validated for each target cardinality.

\item \textbf{Static data only.} The framework assumes the dataset is fixed at outsourcing time. Streaming or append-only datasets would require incremental tree updates and incremental hash maintenance, neither of which is currently supported.

\item \textbf{Decision tree partitioning only.} The ADT authentication mechanism is specific to tree-based anonymization. Other anonymization strategies (e.g., clustering-based approaches like MDAV, or graph-based methods for social network data) would require different authenticated data structures.

\item \textbf{Seven datasets, four cloud profiles.} The evaluation covers 28 scenarios across seven domains and three attack strategies (data dropping, algorithm substitution with a fake hash, and algorithm substitution with a valid hash). The adversary space is still not exhaustive. Section~\ref{sec:informed_attacker} discusses the Informed Attacker ($\mathcal{A}_4$) as a stronger adversary that is analysed theoretically but not yet evaluated empirically. In particular, an adversary that uses a target-aware but approximate algorithm (e.g., a shallower tree or a greedy split that skips expensive features) would produce a valid hash, retain all records, and potentially preserve SHAP distributions within $\varepsilon$. Evaluating this class of adversary is a natural extension of the current work.

\item \textbf{Subsample evaluation.} Experiment~1 uses 8,000-row subsamples and Experiment~3 uses 5,000 rows, not the full dataset sizes. An end-to-end benchmark on synthetic datasets (Section~\ref{sec:scalability}) confirmed that client-side verification scales linearly to $n = 10^6$ rows, but the detection accuracy experiments (Experiment~1) have not been repeated at full dataset scale.

\item \textbf{Single-seed adversary.} The Dumb and Approximate clouds use one fixed random seed for blind splitting, so the per-feature Wasserstein distances reflect a single draw. Characterizing how Layer~3's detection margin varies across seeds is left to future work.
\end{enumerate}

\subsection{Absence of Direct Baseline Comparison}
\label{sec:no_baseline}

No existing published framework simultaneously performs deterministic, probabilistic, and utility-based verification of outsourced data anonymization. This makes a direct apples-to-apples comparison against a competing system impossible.

The closest baselines are:

\begin{enumerate}
\item \textbf{Trap-only verification} (canary records in database auditing). These embed known records into the outsourced data and check whether they survive processing. This is analogous to VeriX-Anon's Layer~2 in isolation. Trap-only methods lack structural verification (no authenticated tree, no hash comparison) and utility verification (no XAI fingerprinting). As demonstrated in Section~\ref{sec:detection}, Layer~2 alone misses the Dumb adversary on every dataset.

\item \textbf{Full re-execution.} The client re-runs the entire Target-Driven anonymization locally. This achieves perfect verification (every aspect of the output can be compared bit-for-bit) but requires $O(n \log n)$ computation to build the decision tree, defeating the purpose of outsourcing. The client must also possess the computational resources, memory, and software stack to execute the full anonymization algorithm. VeriX-Anon's verification cost is $O(n/k)$ for hash traversal plus $O(1)$ for XAI, which is strictly cheaper and does not require the client to build a tree.
\end{enumerate}

Three lines of general verifiable computation are worth comparing against directly, even though none targets anonymization. A zkSNARK such as Spartan~(\cite{setty2020spartan}) gives a succinct proof with no trusted setup, but prover cost scales with circuit size: at $2^{20}$ constraints it needs tens of seconds and gigabytes of prover memory, and a $10^5$-node anonymization tree over $10^6$ rows exceeds that budget, so proving costs more than the original computation. Secure multiparty computation such as SecureML~(\cite{mohassel2017secureml}) keeps data private across two non-colluding servers, but it targets model training rather than integrity verification, assumes no collusion, and moves tens to hundreds of gigabytes offline. Trusted execution such as VC3~(\cite{schuster2015vc3}) runs at near-native speed (4.5 to 8\% overhead) but relies on SGX hardware trust, is bounded by a 512\,MB enclave, and puts side channels out of scope. VeriX-Anon sits at a different point: $O(n/k) + O(1)$ client cost, sub-second at $10^6$ rows, no trusted setup and no special hardware, and it is the only one of the four that verifies utility rather than only structure.

Table~\ref{tab:detection_full} also supports a comparison against isolated verification strategies. Under the calibrated threshold a trap-only approach (Layers~2a and~2b) catches every Lazy cloud via twins but misses every Dumb and Approximate cloud, which retain all records, for 14 of 28. A hash-only approach (Layer~1) catches every Dumb cloud but misses every Lazy and Approximate cloud, also 14 of 28. XAI alone reaches 23 of 28. A sampling audit of a random 10\% of records detects structural changes only probabilistically and still cannot verify utility. VeriX-Anon combines all three mechanisms for 27 of 28 (96.4\%) without forcing the client to build a single tree. Within that combination, Layer~3 is the sole detector for six of the 28 scenarios, the Approximate cloud on every dataset except Nomao, where Layers~1 and~2 all pass and only the XAI check fires; removing Layer~3 would drop the framework from 27 of 28 to 21 of 28. Table~\ref{tab:baseline_comparison} summarises these rates.

\begin{table}[h]
\centering
\caption{Detection accuracy of isolated verification approaches versus VeriX-Anon across 28 scenarios (7 datasets $\times$ 4 cloud profiles), under the per-dataset calibrated threshold. Combining Layers~1 and~2 reaches 21 of 28; Layer~3 adds the six Approximate detections that no other layer can make, taking the framework to 27 of 28.}
\label{tab:baseline_comparison}
\begin{tabular}{|l|c|c|}
\hline
\textbf{Approach} & \textbf{Correct} & \textbf{Accuracy} \\
\hline
Hash-only (Layer 1 alone) & 14/28 & 50.0\% \\
\hline
Trap-only (Layers 2a+2b alone) & 14/28 & 50.0\% \\
\hline
XAI-only (Layer 3 alone) & 23/28 & 82.1\% \\
\hline
Full re-execution & 28/28 & 100\% \\
\hline
Layers 1+2 combined (no XAI) & 21/28 & 75.0\% \\
\hline
\textbf{+ Layer 3 = VeriX-Anon (Proposed)} & \textbf{27/28} & \textbf{96.4\%} \\
\hline
\end{tabular}
\end{table}

VeriX-Anon occupies the space between these two extremes: it provides stronger guarantees than trap-only methods (adding deterministic hash verification and utility-based XAI checks) while remaining orders of magnitude cheaper than full re-execution. As the field of verifiable outsourced anonymization matures, future work should benchmark VeriX-Anon against emerging systems that address the same verification problem.

\subsection{Deployment Considerations}
\label{sec:deployment}
Several practical considerations arise when deploying VeriX-Anon in a real cloud environment. First, the client must retain sufficient local resources to train a Random Forest on 10\% of the data and an XGBoost model on 2,000 rows. On commodity hardware (e.g., a laptop with 8\,GB RAM), both operations complete in under 5 seconds for the dataset sizes tested. For organisations with extremely limited compute (e.g., embedded devices), these operations could be offloaded to a trusted local server.
 
Second, the protocol assumes the cloud returns the full anonymised dataset, the leaf assignment mapping, and the Merkle root hash. The verification-specific communication overhead is analysed in Section~\ref{sec:scalability} and totals approximately 88\,MB for a million-row dataset, modest relative to the dataset itself.
 
Third, returning the tree structure (split features, split values, leaf bounds) to the client does not introduce a privacy risk beyond what the anonymised dataset already reveals. The tree encodes the generalisation logic applied to the data; the client, as the data owner, is entitled to this information and needs it to verify correctness. In a contractual cloud setting, the tree structure would be part of the Service Level Agreement (SLA) deliverables.
 
Finally, integration with real cloud APIs (e.g., AWS Lambda, Azure Functions) would require packaging the ADT construction and Merkle hashing as a cloud-side module and the verification engine as a client-side library. The current Kaggle notebook implementation is a research prototype; production hardening (error handling, streaming hash computation, API wrappers) is engineering work rather than a research contribution and is left to future deployment.

From a managerial perspective, VeriX-Anon changes the cost-benefit calculus of outsourced anonymization auditing. A Data Protection Officer at a hospital or financial institution currently has two options: trust the cloud provider's output without verification, or re-execute the entire anonymization locally (defeating the purpose of outsourcing). VeriX-Anon introduces a third option: spend under one second of local compute to audit the cloud's output across all three verification dimensions. Given that GDPR fines for inadequate anonymization have exceeded EUR~5.88 billion in aggregate since 2018~(\cite{dlapiper2025gdpr}) and that a single outsourcing breach (Capita plc) cost GBP~14 million~(\cite{ico2025capita}), the verification overhead is negligible relative to the compliance risk it mitigates. In practice, VeriX-Anon could be integrated into existing Service Level Agreements as a contractual verification clause, where the cloud provider is required to return the tree structure and Merkle root hash alongside the anonymized data, and the client runs the four-layer audit before accepting delivery.

\section{CONCLUSION}
\label{sec:conclusion}
VeriX-Anon demonstrates that outsourced k-anonymization can be verified without re-executing the anonymization algorithm, using an integrated intelligent auditing system that combines cryptographic, probabilistic, and AI-driven verification. The multi-layered design (Merkle hashing, probabilistic traps via Boundary Sentinels and Twins, SHAP fingerprinting) achieves what no single mechanism can: correct detection in 27 of 28 scenarios spanning seven datasets and four cloud profiles under per-dataset threshold calibration. The single evasion is the Approximate adversary on Nomao, a high-dimensional dataset where honest generalization shifts the SHAP distribution more than the attack does, which marks the operating boundary of utility fingerprinting rather than a systemic design flaw.

The practical implications extend beyond the specific threat models tested here. Any organisation that outsources privacy-sensitive data transformations faces the same verification gap: the inability to confirm, after the fact, that the contracted algorithm was faithfully executed. VeriX-Anon provides a concrete, sub-second audit mechanism that requires only a fraction of the computational resources needed for re-execution ($O(n/k) + O(1)$ vs. $O(n \log n)$). In other words, for a dataset of one million entries at $k = 5$, this equates to 0.788\,s of client-side computation versus rebuilding the decision tree from scratch.

The limitations point to specific next steps. The Nomao miss, where the honest and adversarial SHAP distributions are inverted, makes the case for a utility-based fingerprint that reads the held-out F1 gap directly rather than a distribution shift. The low sentinel counts on imbalanced datasets (4 for Diabetes, 13 for Bank Marketing, with evasion probability up to 0.82) call for boundary-aware injection strategies that widen the probability band when boundary candidates are scarce. Extending the framework to multi-class targets, regression tasks, and streaming data would make it more applicable to the full range of outsourced analytics tasks. Finally, testing the framework against a real-world cloud provider under contractually agreed upon SLA constraints is the ultimate validation.

\appendix
\section{CONFIGURATION PARAMETERS}
\label{app:config}

Table~\ref{tab:full_config} lists all configuration parameters used throughout the experimental evaluation. These values are fixed across all datasets and adversary profiles unless stated otherwise in the main text.

\begin{table}[!h]
\centering
\caption{Complete configuration parameters for VeriX-Anon.}
\label{tab:full_config}
\resizebox{\columnwidth}{!}{
\begin{tabular}{|l|l|}
\hline
\textbf{Parameter} & \textbf{Value} \\
\hline
$k$-anonymity parameter & 5 \\
\hline
Wasserstein threshold $\varepsilon$ & 0.45 \\
\hline
SHAP subsample size & 2,000 rows \\
\hline
Sentinel injection ratio & 2\% of $N$ \\
\hline
Twin injection ratio & 5\% of $N$ \\
\hline
Sentinel boundary band & $P \in [0.45, 0.55]$ \\
\hline
Sentinel perturbation scale & $0.05 \cdot \sigma_j$ per column \\
\hline
Calibration margin ($\varepsilon_d = 1.1\,W_{\max}^{\text{honest}}$) & 1.1 \\
\hline
Random Forest (sentinel gen.) & 50 trees, max depth 5 \\
\hline
XGBoost (XAI fingerprint) & 100 estimators, max depth 6, LR 0.1 \\
\hline
ADT max tree depth & 50 \\
\hline
ADT min leaf size & $2k$ (= 10 at $k = 5$) \\
\hline
Bootstrap resamples (CIs) & 10,000 \\
\hline
Lazy adversary drop fraction $\delta$ & 0.05 \\
\hline
$k$-sweep values (Experiment 3) & $\{2, 3, 4, 5, 7, 10, 12, 15, 20, 25, 30\}$ \\
\hline
Random seed & 42 \\
\hline
\end{tabular}
}
\end{table}

\section{BANK MARKETING COLUMN MAPPING}
\label{app:bank_mapping}

OpenML dataset (ID 1461) encodes Bank Marketing features as V1–V16 instead of the original UCI names~\cite{moro2011bank}. Table~\ref{tab:bank_mapping} shows the mapping used in this paper, with top-3 SHAP features highlighted.

\begin{table}[!h]
\centering
\caption{Bank Marketing (OpenML id=1461) feature name mapping. Top-3 SHAP features marked with $\star$.}
\label{tab:bank_mapping}
\begin{tabular}{|l|l|l|}
\hline
\textbf{OpenML Code} & \textbf{UCI Name} & \textbf{Type} \\
\hline
V1 & age $\star$ & numeric \\
\hline
V2 & job & categorical \\
\hline
V3 & marital & categorical \\
\hline
V4 & education & categorical \\
\hline
V5 & default & binary \\
\hline
V6 & balance $\star$ & numeric \\
\hline
V7 & housing & binary \\
\hline
V8 & loan & binary \\
\hline
V9 & contact & categorical \\
\hline
V10 & day & numeric \\
\hline
V11 & month & categorical \\
\hline
V12 & duration $\star$ & numeric \\
\hline
V13 & campaign & numeric \\
\hline
V14 & pdays & numeric \\
\hline
V15 & previous & numeric \\
\hline
V16 & poutcome & categorical \\
\hline
Class & y (subscribed) & binary target \\
\hline
\end{tabular}
\end{table}

\section{REPRODUCIBILITY CHECKSUMS}
\label{app:checksums}
Table~\ref{tab:hashes} reports the first 8 hex characters of the Merkle root hashes produced by the honest and dumb cloud for each dataset at $k = 5$. These values are deterministic given the same input data, random seed, and configuration. They can be reproduced by running the Kaggle notebook, which is available from the corresponding author upon reasonable request.

\begin{table}[!t]
\centering
\caption{Merkle root hash prefixes (first 8 hex characters) for reproducibility verification. The Approximate cloud produces a valid hash distinct from both Honest and Dumb, confirming that Layer~1 cannot distinguish algorithm substitution when the hash is correctly computed.}
\label{tab:hashes}

\setlength{\tabcolsep}{4pt}
\renewcommand{\arraystretch}{0.95}

\begin{tabular}{|l|c|c|c|}
\hline
\textbf{Dataset} & \textbf{Honest} & \textbf{Dumb} & \textbf{Approx.} \\
\hline
Adult Income & \texttt{4ede8cde} & \texttt{d0592611} & \texttt{ff2b863d} \\ \hline
Bank Marketing & \texttt{36ad5073} & \texttt{cd14b532} & \texttt{bb345a1a} \\ \hline
Diabetes 130-US & \texttt{35b93103} & \texttt{0a6aa553} & \texttt{9eb529ab} \\ \hline
Electricity & \texttt{3f02c07c} & \texttt{72334184} & \texttt{ed70bb02} \\ \hline
Nomao & \texttt{c8cfc526} & \texttt{8f0510d8} & \texttt{bc672297} \\ \hline
Credit Default & \texttt{f6b74d97} & \texttt{c154fd6a} & \texttt{7d38e7f8} \\ \hline
MagicTelescope & \texttt{37643233} & \texttt{711c6634} & \texttt{f524c36e} \\ \hline
\end{tabular}
\end{table}

\section{FULL DETECTION MATRIX}
\label{app:detection}
\begin{figure*}[!h]
\centering
\includegraphics[width=\textwidth]{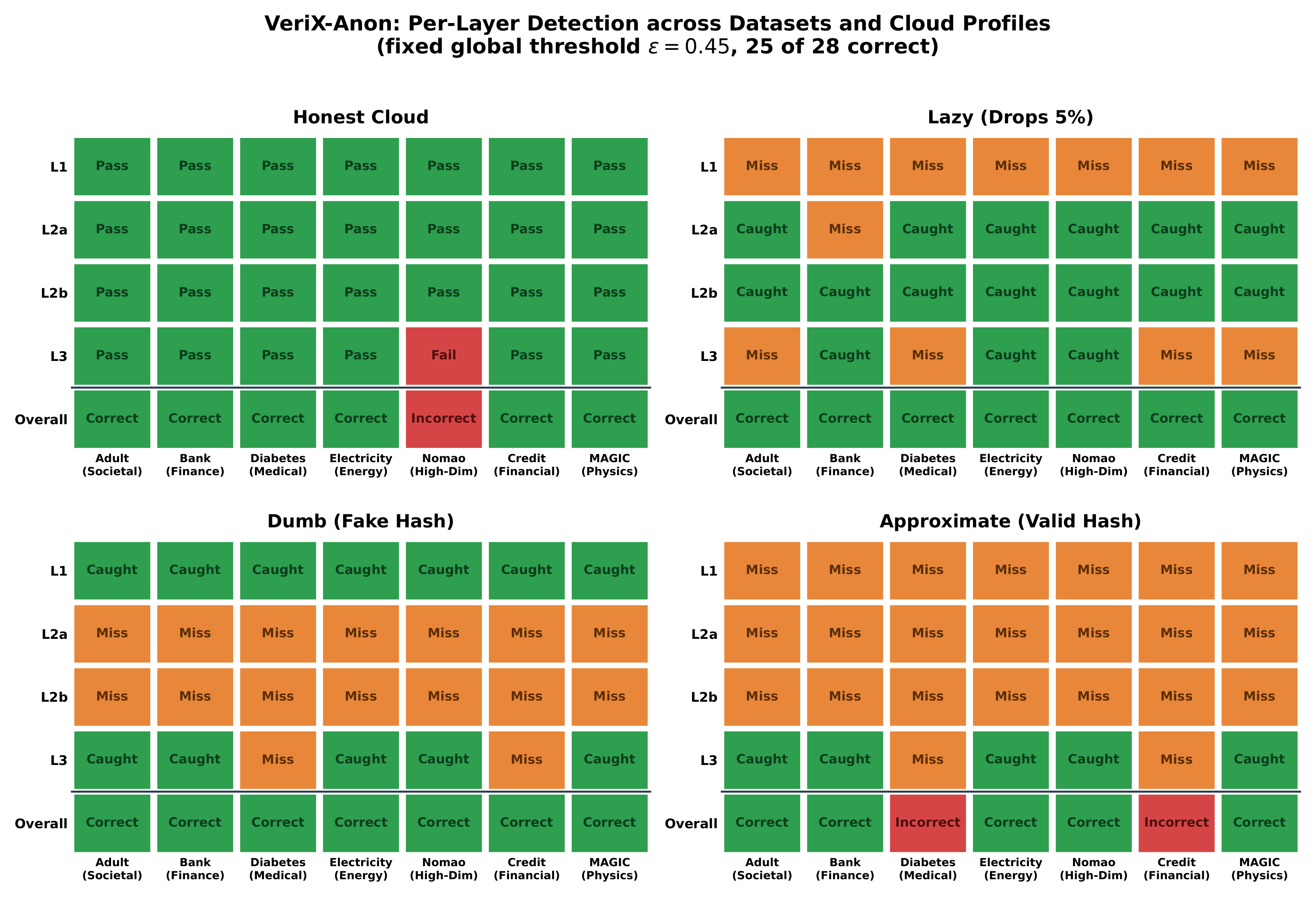}
\caption{Per-layer coverage under the fixed global threshold $\varepsilon = 0.45$ (25 of 28 correct), the companion to the calibrated view in Figure~\ref{fig:heatmap}. Three cells change relative to calibration: honest Nomao becomes a false positive and the Diabetes and Credit Approximate clouds are missed, while Nomao's Approximate cloud is caught. Per-dataset calibration corrects the first three at the cost of the last, moving the framework from 25 of 28 to 27 of 28.}
\label{fig:heatmap_global}
\end{figure*}

Table~\ref{tab:detection_full} gives the per-scenario, per-layer verdict for all 28 scenarios under the per-dataset calibrated threshold. A check mark means the layer behaves correctly (passes an honest cloud or catches a malicious one); a cross means it does not. The only incorrect overall verdict is the Approximate cloud on Nomao. Under the fixed global threshold three cells differ: honest Nomao becomes a false positive, and the Diabetes and Credit Approximate clouds are missed, giving 25 of 28 rather than 27 of 28, shown in Figure~\ref{fig:heatmap_global}.

\begin{table}[!h]
\centering
\caption{Full per-layer detection matrix over 28 scenarios under the per-dataset calibrated threshold. L3 is the XAI layer at the calibrated $\varepsilon_d$. The Approximate cloud reuses the Dumb cloud's blind tree.}
\label{tab:detection_full}
\resizebox{\columnwidth}{!}{
\begin{tabular}{|l|l|c|c|c|c|c|}
\hline
\textbf{Dataset} & \textbf{Profile} & \textbf{L1} & \textbf{L2a} & \textbf{L2b} & \textbf{L3} & \textbf{Overall} \\
\hline
Adult & Honest & \checkmark & \checkmark & \checkmark & \checkmark & \checkmark \\ \hline
Adult & Lazy & $\times$ & \checkmark & \checkmark & \checkmark & \checkmark \\ \hline
Adult & Dumb & \checkmark & $\times$ & $\times$ & \checkmark & \checkmark \\ \hline
Adult & Approx. & $\times$ & $\times$ & $\times$ & \checkmark & \checkmark \\ \hline
Bank & Honest & \checkmark & \checkmark & \checkmark & \checkmark & \checkmark \\ \hline
Bank & Lazy & $\times$ & $\times$ & \checkmark & \checkmark & \checkmark \\ \hline
Bank & Dumb & \checkmark & $\times$ & $\times$ & \checkmark & \checkmark \\ \hline
Bank & Approx. & $\times$ & $\times$ & $\times$ & \checkmark & \checkmark \\ \hline
Credit & Honest & \checkmark & \checkmark & \checkmark & \checkmark & \checkmark \\ \hline
Credit & Lazy & $\times$ & \checkmark & \checkmark & $\times$ & \checkmark \\ \hline
Credit & Dumb & \checkmark & $\times$ & $\times$ & \checkmark & \checkmark \\ \hline
Credit & Approx. & $\times$ & $\times$ & $\times$ & \checkmark & \checkmark \\ \hline
Diabetes & Honest & \checkmark & \checkmark & \checkmark & \checkmark & \checkmark \\ \hline
Diabetes & Lazy & $\times$ & \checkmark & \checkmark & \checkmark & \checkmark \\ \hline
Diabetes & Dumb & \checkmark & $\times$ & $\times$ & \checkmark & \checkmark \\ \hline
Diabetes & Approx. & $\times$ & $\times$ & $\times$ & \checkmark & \checkmark \\ \hline
Electricity & Honest & \checkmark & \checkmark & \checkmark & \checkmark & \checkmark \\ \hline
Electricity & Lazy & $\times$ & \checkmark & \checkmark & \checkmark & \checkmark \\ \hline
Electricity & Dumb & \checkmark & $\times$ & $\times$ & \checkmark & \checkmark \\ \hline
Electricity & Approx. & $\times$ & $\times$ & $\times$ & \checkmark & \checkmark \\ \hline
Magic & Honest & \checkmark & \checkmark & \checkmark & \checkmark & \checkmark \\ \hline
Magic & Lazy & $\times$ & \checkmark & \checkmark & $\times$ & \checkmark \\ \hline
Magic & Dumb & \checkmark & $\times$ & $\times$ & \checkmark & \checkmark \\ \hline
Magic & Approx. & $\times$ & $\times$ & $\times$ & \checkmark & \checkmark \\ \hline
Nomao & Honest & \checkmark & \checkmark & \checkmark & \checkmark & \checkmark \\ \hline
Nomao & Lazy & $\times$ & \checkmark & \checkmark & $\times$ & \checkmark \\ \hline
Nomao & Dumb & \checkmark & $\times$ & $\times$ & $\times$ & \checkmark \\ \hline
Nomao & Approx. & $\times$ & $\times$ & $\times$ & $\times$ & $\times$ \\ \hline
\end{tabular}
}
\end{table}

\section{FULL PER-FEATURE WASSERSTEIN DISTANCES}
\label{app:wd}
Table~\ref{tab:per_feature_full} lists the per-feature SHAP Wasserstein distance for the top-3 features of each dataset under the Honest, Lazy, and Dumb clouds. The Approximate cloud produces the same distances as the Dumb cloud, since the two share the blind tree.

\begin{table}[!h]
\centering
\caption{Per-feature SHAP Wasserstein distance for the top-3 features of each dataset. Dumb and Approximate share a column. Bank features are shown under their UCI names.}
\label{tab:per_feature_full}
\resizebox{\columnwidth}{!}{
\begin{tabular}{|l|l|c|c|c|}
\hline
\textbf{Dataset} & \textbf{Feature} & \textbf{Honest} & \textbf{Lazy} & \textbf{Dumb/Approx.} \\
\hline
Adult & \texttt{age} & 0.200 & 0.300 & 0.512 \\ \hline
 & \texttt{relationship} & 0.088 & 0.159 & 0.210 \\ \hline
 & \texttt{education-num} & 0.143 & 0.228 & 0.266 \\ \hline
\hline
Bank & \texttt{duration} & 0.418 & 0.621 & 1.262 \\ \hline
 & \texttt{age} & 0.191 & 0.231 & 0.172 \\ \hline
 & \texttt{balance} & 0.185 & 0.286 & 0.049 \\ \hline
\hline
Diabetes & \texttt{number\_diagnoses} & 0.087 & 0.181 & 0.242 \\ \hline
 & \texttt{num\_lab\_procedures} & 0.087 & 0.102 & 0.068 \\ \hline
 & \texttt{num\_medications} & 0.122 & 0.095 & 0.086 \\ \hline
\hline
Electricity & \texttt{nswprice} & 0.420 & 0.498 & 0.467 \\ \hline
 & \texttt{date} & 0.265 & 0.334 & 0.426 \\ \hline
 & \texttt{period} & 0.227 & 0.051 & 0.065 \\ \hline
\hline
Nomao & \texttt{V6} & 1.163 & 0.452 & 1.005 \\ \hline
 & \texttt{V97} & 1.108 & 0.752 & 0.888 \\ \hline
 & \texttt{V90} & 0.765 & 0.748 & 0.652 \\ \hline
\hline
Credit & \texttt{x6} & 0.248 & 0.218 & 0.381 \\ \hline
 & \texttt{x12} & 0.119 & 0.194 & 0.188 \\ \hline
 & \texttt{x1} & 0.080 & 0.185 & 0.219 \\ \hline
\hline
Magic & \texttt{fAlpha} & 0.194 & 0.182 & 0.767 \\ \hline
 & \texttt{fSize} & 0.106 & 0.200 & 0.409 \\ \hline
 & \texttt{fLength} & 0.140 & 0.175 & 0.224 \\ \hline
\hline
\end{tabular}
}
\end{table}

\section*{Data Availability Statement}

The datasets analyzed during the current study are publicly available.

The Adult Income dataset is available from OpenML (ID: 1590) at \url{https://www.openml.org/d/1590}.

The Bank Marketing dataset is available from OpenML (ID: 1461) at \url{https://www.openml.org/d/1461}.

The Diabetes 130-US Hospitals dataset is available from the UCI Machine Learning Repository at \url{https://archive.ics.uci.edu/dataset/296/diabetes+130-us+hospitals+for+years+1999-2008}.

The Electricity dataset is available from OpenML (ID: 151) at \url{https://www.openml.org/d/151}.

The Nomao dataset is available from OpenML (ID: 1486) at \url{https://www.openml.org/d/1486}.

The Credit Default (Taiwan) dataset is available from OpenML (ID: 42477) at \url{https://www.openml.org/d/42477}.

The MagicTelescope dataset is available from OpenML (ID: 1120) at \url{https://www.openml.org/d/1120}.

The code used in this study is available from the corresponding author upon reasonable request.

\printcredits

\bibliographystyle{cas-model2-names}

\bibliography{cas-refs}
\newpage

\end{document}